\documentclass[preprint,nonumbib]{ptephy}

\preprintnumber{YGHP-16-02} 

\usepackage{amsmath}
\usepackage{graphicx}
\usepackage[utf8]{inputenc}
\usepackage{amssymb}
\usepackage{amsfonts}
\usepackage{amsbsy}
\usepackage{url}
\usepackage{enumerate}

\newcommand{\refer}[1]{(\ref{#1})}
\newcommand{\abs}[1]{\left|#1\right|}
\newcommand{\mean}[1]{\left\langle #1\right\rangle}
\newcommand{\Exp}[1]{\mathrm{e}^{#1}}
\newcommand{\arccosh}{\mbox{arcosh}}

\begin{document}

\title{Exact BPS domain walls at finite gauge coupling}

\author{\name{Filip Blaschke}{1}}

\address{\affil{1}{Department of Physics, Yamagata University, Kojirakawa-machi 1-4-12, Yamagata,
Yamagata 990-8560, Japan}
\email{fblasch(at)sci.kj.yamagata-u.ac.jp}}

\begin{abstract}%
Bogomol'nyi-Prasad-Sommerfield solitons in models with spontaneously broken gauge symmetry have been intensively studied at infinite gauge coupling limit, where
the governing equation --  so-called master equation -- is exactly solvable. Except for a handful of special solutions, the standing impression is that analytic results at finite coupling are generally unavailable. The aim of this paper is to demonstrate, using domain walls in Abelian-Higgs models as the simplest example, that exact solitons at finite gauge coupling can be readily obtained if the number of Higgs fields ($N_F$) is large enough.
In particular, we present a family of exact solutions, describing $N$ domain walls at arbitrary positions in models with at least $N_F \geq 2N+1$. We have also found that adding together any pair of solution can produce a new exact solution if the combined tension is below a certain limit.
\end{abstract}


\maketitle

\section{Introduction}

Domain walls are planar-like objects, which separates two distinct phases/states of matter. Perhaps the  most iconic examples are walls in ferromagnets, where `domains' of uniform orientation of magnetic moments are punctuated by thin layers, within which moments change rapidly to accommodate the relative difference. Numerous other examples can be found in solid-state physics.
The concept, however, found its place in various others fields. In theoretical physics, domain walls are often used as low energy toy models of both D-branes \cite{Portugues, Tong2} in string theory and `branes' in the brane world scenario \cite{Rubakov}. The lowest common denominator in all their incarnations, however, is the existence of distinct phases/states of matter,  which makes domain walls prototypes of topological solitons \cite{Manton}.

To model a domain wall in field theory, in principle all that is required are discrete and degenerate vacua, which domain wall separates. 
The simplest example is a scalar field theory with the double-well potential, the well known $\mathbb{Z}_2$ kink \cite{Vachaspati}. 
However, in generic scalar theories, a configuration of more than a one domain wall cannot be static, due to attractive scalar forces. Dynamical multi-soliton configurations are therefore difficult to study in such theories, other than via numerical or perturbative methods. Only in very special cases, we are able to write-down exact dynamical multi-soliton solutions, such as in the famous Sine-Gordon model. 

Hence, to understand multi-domain walls (and many other topological solitons as well), without the need of dealing with dynamics, it is advantageous to use 
supersymmetric (SUSY) gauge theories.
In these models, attractive scalar forces between solitons can be exactly compensated by repulsive massive vector interactions (originating from spontaneously broken gauge symmetry). Thus, arbitrary configurations of solitons  can be static.
Solitons with such property are called Bogomol'nyi-Prasad-Sommerfield (BPS) solitons. They can be shown to possess the least amount of energy in their topological sector (so-called Bogomol'nyi bound \cite{Bog}) and they also partially preserve SUSY. 

BPS solitons have a further advantage. Instead of dealing with complicated second-order equations of motion, BPS solitons can be shown to satisfy a much simpler set of first-order equations, correspondingly called the BPS equations.  
These equations have many interesting properties and
attract scientific interest on their own.

If the gauge symmetry is only partially broken by scalar fields -- the so-called Coulomb phase -- prototypal solitons are magnetic monopoles \cite{tHooft, Polyakov} and to them related instantons  \cite{Belavin}. Their BPS equations can be solved exactly via the famous Nahm equations \cite{Nahm} and ADHM construction \cite{Atiyah}, respectively, providing an analytic form for static multi-monopole and multi-instanton configurations. This reflects the underlying integrability of BPS equations in the Coulomb phase. 

If the gauge symmetry is fully broken by scalar fields -- the so-called Higgs phase -- the prototype solitons are vortices \cite{Abrikosov, Nielsen, Hanany, Auzzi} and domain walls \cite{Tong1, Sakai2}. In contrast with the Coulomb phase, similar  analytic techniques for solving BPS equations in the Higgs phase are not available, as they are not integrable \cite{Nitta}.
Although isolated exact solutions were found for domain walls \cite{Sakai3, Sakai4}, they only seem to prove the general rule.
 
This paper is dedicated to extending the knowledge about exact solutions for solitons in the Higgs phase. In particular, we concentrate on domain walls, but our results have obvious implications to other BPS solitons in the Higgs phase.

We consider a part of the bosonic sector of Abelian $N=2$ SUSY gauge theory with $N_F$ Higgs fields. Such model has $N_F$ discrete vacua and consequently supports configurations of up to $N_F-1$ elementary domain walls. We present the model in detail in Sec.~\ref{sec:model}, where we also discuss properties of walls and of multi-wall configurations at some length.

To construct BPS domain walls we employ the so-called moduli matrix approach (for a review see \cite{Eto1}).
This technique involves a change of variables which solves all BPS equations identically, except one. The remaining equation called \emph{the master equation}, has the following general form
\begin{equation}\label{eq:masterw0}
\frac{1}{\tilde g^2}\partial_x^2 u = 1- \Omega_0(x)e^{-u}\,,
\end{equation}
where $\tilde g$ is proportional to the gauge coupling constant and $u$ is the moduli field related to Higgs fields and gauge fields via the moduli matrix approach. 
We call $\Omega_0(x)$ the `source term' since it specifies number, mass and positions of walls. In fact, together with the boundary condition $u\to\log\Omega_0(x)$ at the spatial infinity, $\Omega_0$ fully determines the solution. 

Not surprisingly, the non-linear second order differential equation \refer{eq:masterw0} is very hard to solve exactly.
Indeed, up to now, there is only a handful of exact solutions for special choices of $\Omega_0$.
Chronologically, the first exact solution was a junction of domain walls reported in \cite{Sakai4}. From this solution, one can extract a single wall solution, which was added to two other exact single wall solutions and one double wall solution in \cite{Sakai3}.
Outside of these important findings, the master equation of the type \refer{eq:masterw0} is believed to be impregnable by analytic means. 

In this paper, we will present many new exact solutions for multi-domain walls.
In particular, we will show that some exact solutions of \refer{eq:masterw0}, each with different $\Omega_0$, can be combined to gain novel solutions, with new $\Omega_0$'s. These solutions, which we call \emph{chains}, can under certain conditions generate an unlimited wealth of exact solutions, which we specify in Sec.~\ref{sec:results}. Furthermore, we will show that beyond chains there is a bewildering maze of exact solutions based upon a family of ansatzes, which we present and study in Sec.~\ref{sec:grinder}. Both of these findings reveal unexpected analytic structure within master equation for domain walls and strongly implies the same for other topological solitons. We will discuss these implications in Sec.~\ref{sec:conclusions}.

\section{Master equation for Abelian domain walls}\label{sec:model}

\subsection{Abelian-Higgs theory}

Let us consider a $U(1)$ gauge theory in (3+1)-dimensions\footnote{The number of spatial dimensions plays no role in our discussions. We keep it to be three for simplicity, but all results presented in subsequent sections are valid in any number of spatial dimensions.} with $N_F$ complex scalar fields assembled into the row vector $H \equiv (H^1,H^2,\ldots ,H^{N_F})$ and a real scalar field $\sigma$. The Lagrangian is given as:
\begin{align}
\label{eq:lag} {\mathcal L} & = -\frac{1}{4 g^2}(F_{\mu\nu})^2+\frac{1}{2g^2}(\partial_{\mu}\sigma)^2+\abs{D_{\mu}H}^2-V\,, \\
V & = \frac{g^2}{2}\bigl(v^2-HH^{\dagger}\bigr)^2+\abs{\sigma H-H M}^2\,,
\end{align}
where $g$ is the gauge coupling constant and the parameter $v$ is the vacuum expectation value of the Higgs field. In the context of supersymmetry, $v^2$ can be identified with the so-called Fayet-Iliopoulous parameter. The mass matrix is given as $M = \mbox{diag}(m_1,\ldots, m_{N_F})$.
Notice that the trace part of $M$ can be changed arbitrarily by shifting $\sigma\to \sigma+k$, $\mbox{Tr}[M]\to \mbox{Tr}[M]-k$, which leaves the Lagrangian unaffected. We will use this freedom to make all entries in $M$ negative (or zero) by setting $k = N_F\mbox{max}\{m_i\}$. The benefit of this choice will be apparent when we discuss positions of domain walls. Also, we can always demand that the masses are ordered $m_1 \geq \ldots \geq m_{N_F}$, without loss of generality. This makes $m_1 = 0$.
  
We adopt following conventions:
\begin{align}
\eta_{\mu\nu} & = \mbox{diag}(+,-,-,-)\,, \\
F_{\mu\nu} & = \partial_{\mu}w_{\nu}-\partial_{\nu}w_{\mu}\,, \\
D_{\mu}H & = \partial_{\mu}H+ iw_{\mu} H\,.
\end{align}

Non-generic values of coupling constants in front of $(\partial \sigma)^2$ and in the potential $V$ allow us to embed this model into a supersymmetric model with eight supercharges by adding appropriate bosonic and fermionic fields.
This `effective' supersymmetry enables to reduce the full equations of motion
by one order. The reduced equations are known as 1/2 BPS equations.

Vacua of  \refer{eq:lag} are constant field configurations with vanishing potential $V = 0$. There are precisely $N_F$ such configurations, which we label as $\mean{1}, \ldots \mean{N_F}$, with the representative values
\begin{equation}
H^{\langle k\rangle} = (0,\ldots , \underbrace{v}_{k-\mbox{th}}, \ldots , 0)\,, \hspace{3mm} \sigma^{\langle k\rangle} = 
m_k\,, \hspace{3mm} (k=1,\ldots , N_F)\,.
\end{equation} 

The Existence of discrete vacua is what makes domains walls possible. To be specific, since there are $N_F$ vacua we expect there to be a domain wall for every transition between a pair of vacua $\mean{i}-\mean{j}$, where $i<j$ (if $i>j$ such transitions are called anti-walls). Thus, there are $\binom{N_F-1}{2}$ different domain walls. Let us denote each transition as $\mean{i\, j}$ for brevity. 
Out of these, only transitions between successive vacua $\mean{i\, i+1}$ are considered as `elementary', while other transitions can be showed as composite configurations of elementary walls, as we will discuss in subsection \ref{sec:master}.
 
\subsection{1/2 BPS equations}

Let us construct 1/2 BPS domain walls along, say, the first direction $x^1$. To that goal we will assume that all fields depend only on $x^1$ coordinate and, as a gauge choice, we take all gauge fields  to be zero $w_\mu=0$.
The BPS equations are found using the well-known technique of bounding the energy density from below due to Bogomol'nyi \cite{Bog}   
\begin{align}
{\mathcal E} & = \frac{1}{2g^2}\bigl(\partial_1 \sigma\bigr)^2+\frac{g^2}{2}\Bigl(v^2-\abs{H}^2\Bigr)^2+\abs{\partial_1 H}^2 + \abs{\sigma H-H M}^2 \nonumber \\ 
 & = \frac{1}{2g^2}\Bigl(\partial_1 \sigma - \eta g^2 \bigl(v^2-\abs{H}^2\bigr)\Bigr)^2+\abs{\partial_1H+\eta \bigl(\sigma H-H M\bigr)}^2+\eta v^2\partial_1\sigma\nonumber \\
 & \label{eq:bpswall} -\eta\partial_1\bigl(\abs{H}^2\sigma -HMH^{\dagger}\bigr)
 \geq \eta {\mathcal T}- \eta\partial_1 {\mathcal J}\,,
\end{align} 
where ${\mathcal T} = v^2\partial_1\sigma$ is energy density of the domain wall and  ${\mathcal J}= \abs{H}^2\sigma -HMH^{\dagger}$ is a boundary term. Thus, the energy per unit area (tension) of the domain wall is never less than $\eta\, T$, where
\begin{equation}\label{eq:topch}
 T =  v^2 \int\limits_{-\infty}^{\infty}d x^1\, \partial_1 \sigma = v^2 \Delta m
\end{equation}
and $\Delta m$ is a difference of masses of respective vacua.\footnote{Notice that due to the ordering of masses this quantity is always positive for walls $\eta =1$ and negative for anti-walls $\eta = -1$. In this way  the combination $\eta\Delta m = \abs{\Delta m}$ is always positive.}
The minimum $E = \eta T$ is achieved if  the following set of first order equations are obeyed
\begin{gather}
\label{eq:bpsw1} \partial_1 H +\eta \bigl(\sigma H-H M\bigr) = 0\,, \\
\label{eq:bpsw2} \partial_1\sigma = \eta g^2\bigl(v^2-\abs{H}^2\bigr)\,.
\end{gather}
Parameter $\eta^2 = 1$ labels whether the solution is either a wall ($\eta = 1$) or an anti-wall ($\eta = -1$). 

BPS equations \refer{eq:bpsw1}-\refer{eq:bpsw2} can be reduced to a single second order partial differential equation by using the moduli matrix approach \cite{Eto1}. First, let us solve Eq.~\refer{eq:bpsw1} identically using the ansatz
\begin{equation}
\label{eq:moduliw} H = v e^{-u/2}H_0e^{\eta M x^1}\,, \hspace{5mm} \eta\, \sigma = \frac{1}{2}\partial_1 u\,, 
\end{equation}
where $u$ is a new field variable and $H_0$ is a row vector of $N_F$ constants, generically called the moduli matrix. Notice that the pair $\{u,\, H_0\}$ does not determine $\{H,\,\sigma \}$ uniquely, since 
the so-called $V$-transformation $\{u,\, H_0\}\to \{u+2c,\, e^cH_0\}$, where $c \in \mathbb{C}$, leave the assignment \refer{eq:moduliw} unaltered. The second BPS equation \refer{eq:bpsw2} turns into the master equation
\begin{equation}
\label{eq:masterw} \frac{1}{2g^2v^2}\partial_1^2 u = 1-\Omega_0e^{-u}\,, \hspace{5mm} \Omega_0= H_0 \Exp{2\eta M x^1}H_0^{\dagger}\,.
\end{equation}

\subsection{The master equation}\label{sec:master}

As an explicit example, let us consider the case $N_F = 2$, $M= \mbox{diag}(0,-m)$. There is only one domain wall $\mean{12}$, which interpolates between the second vacuum $\sigma \to -m$ at $x^1\to -\infty$ and the first vacuum $\sigma \to 0$ at $x^1\to \infty$. Its tension is $T_{\mean{1\, 2}} = v^2 m$. The generic moduli matrix can be written as $H_0 = (1,e^{m R})$ by fixing the first entry via the $V$-transformation. Parameter $R$ represents the position of the wall and we can set $R=0$ by translation symmetry. The master equation \refer{eq:masterw}
in such a case reduces to ($x^1 \equiv x$, $\eta =1$)
\begin{equation}\label{eq:masterw2}
\frac{1}{2g^2v^2}\partial_{x}^2u = 1-\Bigl(1+e^{ -2m x}\Bigr)e^{-u}\,.
\end{equation}
The general solution of this equation is unknown, although three exact solutions for specific values of the ratio $m/(gv)$ have been found \cite{Sakai3}. We will discuss these solutions in the next section. 

\begin{figure}[!h]
\includegraphics[width = 0.7\textwidth]{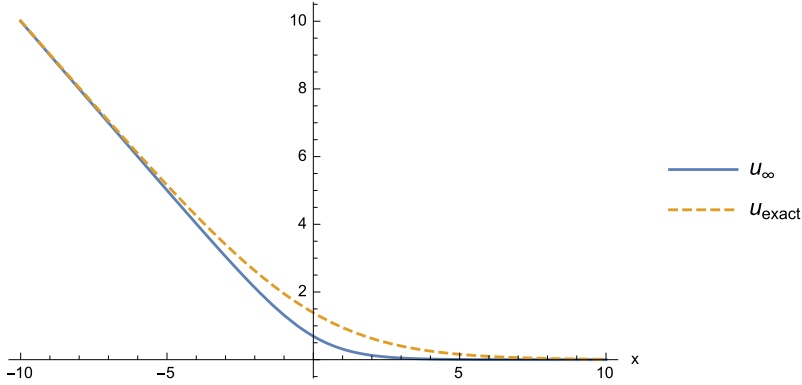}
\caption{\small Comparison
between approximate $u_{\infty}$ solution (solid blue line) and exact $u_{\mbox{\footnotesize exact}}$ solution (dashed yellow line) to the master equation \refer{eq:masterw2}. The values of parameters are $m=1$ and $2g^2v^2 =1$.}
\label{fig:wall}
\end{figure}

Apart from these solutions (or solving the master equation numerically), one can develop a qualitative feeling about the shape of the solution via the so-called infinite gauge coupling limit  $g\to\infty$. Notice that in this limit the left-hand side of \refer{eq:masterw2} vanishes and right-hand side can be solved algebraically as
\begin{equation}
u_{\infty} = \log\Bigl(1+e^{-2m x}\Bigr)\,.
\end{equation}
The difference between an exact solution and $u_{\infty}$ is generally confined to a finite region around domain wall as illustrated in Fig.~\ref{fig:wall}. Thus, the general shape of the solution is faithfully represented by $u_{\infty}$.
This is also the reason why the infinite gauge coupling limit is used almost exclusively in studies about BPS domain walls, or other solitons in the Higgs phase.

Let us now consider the case $N_F = 3$ with mass matrix $M=\mbox{diag}(0,-m,-2m)$.
It is easy to see that there are two elementary walls representing the transitions $\mean{1\, 2}$ and $\mean{2\, 3}$ given by moduli matrices $H_0 = (1,1,0)$ and $H_0 = (0,1,1)$, respectively. Indeed, these choices would produce essentially identical master equations as in Eq.~\refer{eq:masterw2}. On the other hand, if we choose $H_0 = (1,e^{R/2},1)$ we  obtain a new master equation 
\begin{equation}\label{eq:example}
\frac{1}{2g^2v^2}\partial_{x}^2u = 1-\Bigl(1+e^{R}e^{-2m x}+e^{-4m x}\Bigr)e^{-u}\,,
\end{equation}
which is the first example of the so-called composite wall. This configuration describes the transition $\mean{1\, 3}$. Depending on the value of the parameter $R$, the configuration on its way from $\mean{1}$ vacuum at $x\to\infty$ to $\mean{3}$ vacuum at $x\to-\infty$ can approach the $\mean{2}$ vacuum arbitrarily close. 
For positive and large $R$ we can interpret the configuration as a pair of elementary walls separated by the distance $\sim R/m$. Indeed, in the vicinity of the point $x = R/(2m)+\tilde x$, where $\abs{\tilde x}\ll R/(2m)$ we have\begin{equation}
\frac{1}{2g^2v^2}\partial_{x}^2u \approx 1-\Bigl(1+e^{-2m \tilde x}+e^{-2 R}\Bigr)e^{-u}\,,
\end{equation}
which is nearly the transition $\mean{1\, 2}$ up to the negligible term $e^{-2R}\ll 1$, while near the point $x = -R/(2m)+\tilde x$ we get
\begin{equation}
\frac{1}{2g^2v^2}\partial_{x}^2u \approx 1-\Bigl(1+e^{2R}\bigl(e^{-2m \tilde x}+1\bigr)\Bigr)e^{-u}\,.
\end{equation}
We can shift the field $u\to u +2R$ to make the first term in the parenthesis negligible, thus approximately obtaining an equation for the elementary wall $\mean{2\, 3}$.

In other words, for sufficiently big $R$ the solution describes a pair of elementary walls located at
\begin{equation}
x_1 = \frac{R}{m}\,, \hspace{5mm} x_2 = -\frac{R}{m}\,.
\end{equation} 
In the region between the walls, the fields $H$ and $\sigma$ are nearly at the second vacuum and in the limit $R\to \infty$, the whole configuration becomes $\mean{2}$. On the other hand, when $R$ is close to zero or negative the two walls merge together: they form so-called `compressed' wall. In the extreme limit $R\to-\infty$ the master equation \refer{eq:masterw3} reduces to \refer{eq:masterw2} and we have pure transition $\mean{1\, 3}$ with a single domain wall.
An exact solution of Eq.~\refer{eq:example} was reported \cite{Sakai3}. Again, we will discuss it in the next section.

Let us now make a few comments about the general case. For ease of reference let us repeat here the master equation \refer{eq:masterw} in a fully unpacked way  ($x^1 \equiv x$, $\eta =1$):
\begin{equation}\label{eq:masterw3}
\frac{1}{2g^2v^2}\partial_x^2 u = 1-e^{-u}\Bigl(\sum\limits_{i=1}^{N_F}h_i^2e^{-2 m_i x}\Bigr)\,.
\end{equation}
where $h_i$ are absolute values of moduli matrix $H_0$ and $m_i$ are diagonal elements of the mass matrix $M$.
Notice that not all moduli parameters and mass parameters are independent. We can always make the following transformation 
\begin{align}
\label{eq:sym1} u(x) &\to u(x+\alpha)+\beta+\gamma x\,, \\
\label{eq:sym2} m_i & \to m_i-\gamma/2\,, \\
\label{eq:sym3} h_i &\to e^{\gamma \alpha/2-m_i \alpha+\beta}\,, 
\end{align}
without changing the equation. This redundancy is a manifestation of $V$-transfor-mation, translation invariance and reparametrization freedom of the mass matrix.

Generalizing what we learned in particular cases, there are $N_F-1$ elementary walls $\mean{i\, i+1}$, which are obtained by setting $h_i = 1$, $h_{i+1}= 1$ for $i=(1,\ldots, N_F-1)$ and $h_j =0$ otherwise. On the other hand, if there are three or more non-zero moduli parameters, we have a configuration of two or more elementary walls. In the maximal case where all moduli are non-zero, we have everything between a system of well separated $N_F-1$ elementary walls and a single domain wall $\mean{1\, N_F}$.

Can we tell which elementary walls are compressed, which are isolated and at what positions walls (compressed or elementary) roughly are just from moduli parameters? Yes. Let us concentrate on $i$-th elementary domain wall.
We can estimate its position by comparing the factors $h_i^2 e^{2 m_i x}$ and $h_{i+i}^2 e^{2 m_{i+i} x}$ on the right-hand side of Eq.~\refer{eq:masterw3}.
If neither of these numbers (at given $x$) is significantly bigger than the remaining factors in Eq.~\refer{eq:masterw3} we are nowhere close to either $\mean{i}$ or $\mean{i+1}$ vacuum and hence we are also far away from $\mean{i\, i+1}$-th wall. On the other hand, if one of these numbers is much bigger then the rest, we are close to the corresponding vacuum, but again not close to the wall. Obviously, both factor must be equally dominant over remaining terms, in order to be close to $i$-th elementary wall. The point where both factors equal is
\begin{equation}\label{eq:exes}
x_i = \log\bigl(h_{i+1}/h_i\bigr)/(m_{i}-m_{i+1})\,.
\end{equation}
But $x_i$ will faithfully represent position of the wall only if the numbers $h_i^2 e^{2 m_i x_i}$ and  $h_{i+i}^2 e^{2 m_{i+i} x_i}$ really are much bigger then all remaining factors. This happens when $i$-th wall is isolated from its neighbours, that is  if $x_{i-1}\ll x_i \ll x_{i+1}$. If for example $x_{i-1}$ is close or bigger than $x_i$ then both elementary walls are compressed with each other and located at the center of mass of the two wall system:
\begin{equation}
\tilde x_i = \frac{(m_i-m_{i+1})x_i+(m_{i-1}-m_i)x_{i-1}}{m_{i-1}-m_{i+1}} = \frac{1}{m_{i-1}-m_{i+1}}\log\Bigl(h_{i+1}/h_{i-1}\Bigr)\,.
\end{equation}
This is so because ordering of masses $m_i\geq m_{i+1}$ implies that vacua, which are visited by the configuration along the $x$-axis, must be ordered as well. Therefore, the elementary domain walls cannot have arbitrary positions and in fact $i$-th wall must be placed on the right side of $(i+1)$-th wall. 

In general, if we work out the values of all $x_i$, $i=1, \ldots, N_F-1$ given in Eq.~\refer{eq:exes} the ordering of these points tell us which walls are isolated and which are compressed. If they are ordered as $x_1\gg\ldots \gg x_{N_F-1}$, there are $N_F-1$ isolated walls located approximately at those points. In the other extreme, if  the center of mass of first $N_F-2$ elementary walls $\tilde x$ is smaller than $x_{N_F-1}$, then there is only one compressed domain wall located at their mutual center of mass. 

From Eq.~\refer{eq:exes} we see that not all $h_i$'s are necessary to describe positions of walls. We can use the symmetry transformation \refer{eq:sym1}-\refer{eq:sym3} to reduce their number by two. Let us fix this freedom and rewrite moduli $h_i$ into the standard form, where their meaning becomes more transparent:
\begin{equation}\label{eq:has}
h_i = \prod\limits_{j=1}^{i-1}e^{R_j}\,, \hspace{3mm} i = (1,\ldots N_F-1)\,.
\end{equation}  
By virtue of this assignment $h_1 = 1$. New parameters $R_i \equiv  x_i (m_i-m_{i+1})$ are weighted positions of the walls. Since the position of the center-of-mass of the entire system is not very interesting moduli, let us fix it at $x=0$. In other words we demand $R_1+\ldots+R_{N_F-1} = 0$. This makes $h_{N_F} =1$. Furthermore, we will fix the remaining symmetry by demanding that $m_1= 0$. We will call a configuration of domain walls with center of mass fixed at origin and $m_1 =0$ a \emph{balanced} configuration. Any initially unbalanced configuration of domain walls, that is $h_1\not = 1$, $h_{N_F}\not = 1$ and $m_1 \not= 0$ can be turned into a balanced one via the following transformation of the solution $u(x)\to u(x+\alpha)+\beta+2m_1 x$, where
\begin{align}\label{eq:balance1}
 \alpha & = \frac{1}{m_1-m_{N_F}}\log\bigl(h_{N_F}/h_1\bigr)\,,\\
 \label{eq:balance2} \beta & = \frac{2m_1}{m_1-m_{N_F}}\log \bigl(h_{N_F}\bigr)-\frac{2m_{N_F}}{m_1-m_{N_F}}\log \bigl(h_1\bigr)\,.
\end{align}

\subsection{Crude model of the domain wall}\label{sec:crude}

The domain wall is known to possess a three-layered structure \cite{Shifman, Eto1}. The outer two layers of the domain wall are customarily called the `skin' and here we call the inner layer  the `core'. The skin is characterized by the rapid decay of the corresponding Higgs field, from the asymptotic value $H = v$ in the vacuum outside of the domain wall to (almost) zero. The core is a region where all Higgs fields remain very close to zero. This means that in the core the gauge symmetry is almost restored, and we can understand the skin as a transitional region from the spontaneously broken phase outside of the domain wall to the unbroken phase in the core.

\begin{figure}[!h]
\includegraphics[width = 0.8\textwidth]{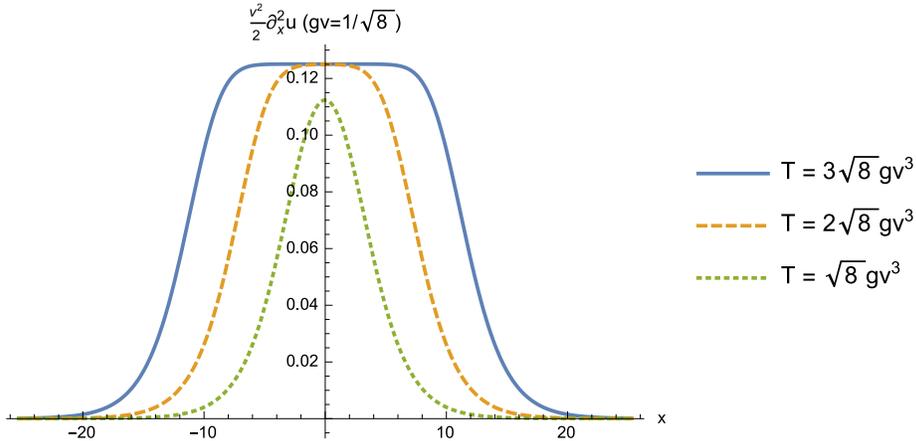}
\caption{\small Numerical solution of the master equation with $\Omega_0 = 1+e^{-2m x}$, illustrating the difference in shape of tension density $v^2 \partial_x^2 u/2$ for various values of total tension $T$. We see that core of the domain wall (the plateau) develops only when total tension is sufficiently big.}
\label{fig:tensions}
\end{figure}

The core, however, is not present for all domain walls, but it develops only when the total tension is sufficiently big. Let us again consider the $N_F =2$ example with the mass matrix $M = \mbox{diag}(0,-m)$ and $H_0 = (1,1)$. On Fig.~\ref{fig:tensions} we show energy density profiles of three domain walls with increasing values of total tension. The three-layered structure is most visible for the heaviest wall, where the plateau in the energy density is clearly present. The lightest wall is visibly core-less and has only skin, while the middle wall is just at the point, where core starts to develop.

The existence of the core can be partially understood from the BPS equations \refer{eq:bpsw1}-\refer{eq:bpsw2}. 
In particular, the second equation \refer{eq:bpsw2} can be seen as a condition that the value of the tension density $v^2\partial_x \sigma$ cannot be at any point higher than $g^2 v^4$ (In Fig.~\ref{fig:tensions} this value is exactly 0.125). Notice that this upper bound can be reached only when Higgs fields are zero, which is the unbroken phase. Hence, if we assume that the (maximal) skin energy is independent of the mass (as we will show is the case) and that the total tension $v^2 \Delta m$ is significantly higher than $g v^3$, the core of the domain wall must develop, in order to store the excess energy. 

This observation implies that the effective radius of the core is linearly proportional to the tension, $R_c \sim \Delta m$. In fact, ignoring skin altogether we easily estimate $R_c$ by comparing energy inside the core $g^2v^4 R_c$ to the total energy $v^2\Delta m$, thus obtaining $R_c \sim \tfrac{\Delta m}{g^2 v^2}$.  

\begin{figure}[!h]
\includegraphics[width = 0.8\textwidth]{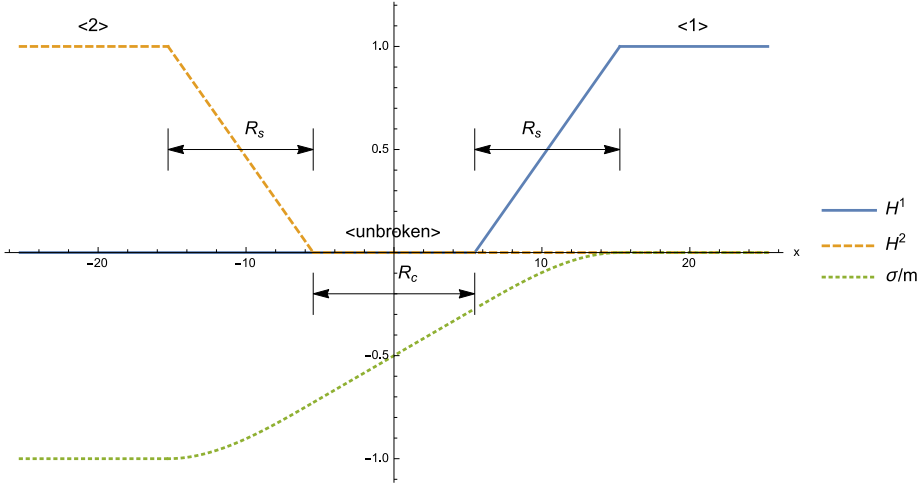}
\caption{\small The crude model of the domain wall, where Higgs fields are pice-wise linear functions.
}
\label{fig:crudemodel}
\end{figure}

We can gain much more insight into the structure of the domain wall by imagining very crude model as depicted on Fig.~\ref{fig:crudemodel}. There, we take the Higgs fields as piece-wise linear functions. Outside the domain wall they are exactly fixed at given vacua ($\mean{1}$ on the right side an $\mean{2}$ on the left side). Inside the skin they linearly fall to zero across the interval of size $R_s$ and remains vanishing in the core of width $R_c$.

The $\sigma$ field is determined by demanding that it is a continuous function and inside every region a solution of the second BPS equation \refer{eq:bpsw2}. 
Carrying out the integration of Eq.~\refer{eq:bpsw2} in each region we find
\begin{align}
\sigma_{\small \mbox{core}} & = g^2v^2 x-\frac{m}{2}\,, \\ 
\sigma_{\small \mbox{right skin}} & = g^2v^2\Bigl(x+\frac{(R_c-2x)^3}{24 R_s^2}\Bigr)-\frac{m}{2}\,, \\
\sigma_{\small \mbox{left skin}} & = g^2v^2\Bigl(x-\frac{(R_c+2x)^3}{24 R_s^2}\Bigr)-\frac{m}{2}\,, \\
\sigma_{\mean{1}} & = 0\,, \\
\sigma_{\mean{2}} & = -m\,.
\end{align}

The integration constants are fixed in the following way. In the core we demand $\sigma_{\small \mbox{core}} (0)= -m/2$  by reflection symmetry. In both skins we determine the integration constants by sewing the $\sigma_{\small \mbox{right skin}}$ and $\sigma_{\small \mbox{left skin}}$ with $\sigma_{\small \mbox{core}}$ at the points $x= \pm R_c/2$. Interestingly, remaining continuity conditions $\sigma_{\small \mbox{right skin}} = \sigma_{\mean{1}}$ and $\sigma_{\small \mbox{left skin}}= \sigma_{\mean{2}} $ at the points $x = \pm R_c/2\pm R_s$ gives a unique relation between $R_s$ and $R_c$ in the form
\begin{equation}\label{eq:relation}
R_c = \frac{m}{g^2v^2}-\frac{4}{3}R_s\,.
\end{equation}
Using this relation we can calculate the tension of the right skin to be 
\begin{align}
T_{\small \mbox{right skin}} & = v^2 \int\limits_{R_c/2}^{R_c/2+R_s}\partial_x \sigma_{\small \mbox{right skin}} \, d x = 
v^2\sigma_{\small \mbox{right skin}}\Big|_{R_c/2}^{R_c/2+R_s}\nonumber \\
& = v^2\bigl(0 -g^2v^2 R_c/2+m/2\bigr) = \frac{v^2m}{2}-\frac{g^2v^4 R_c}{2}\,.
\end{align}
By symmetry we have $T_{\small \mbox{skin}} = 2 T_{\small \mbox{right skin}}  = v^2m-g^2v^2 R_c$.
Combining this with the tension of the core $T_{\small \mbox{core}} = g^2v^2 R_c$ we see that in total we have 
$T = T_{\small \mbox{skin}} +T_{\small \mbox{core}} = v^2 m$, as it should be.

\begin{figure}[!h]
\includegraphics[width = 0.8\textwidth]{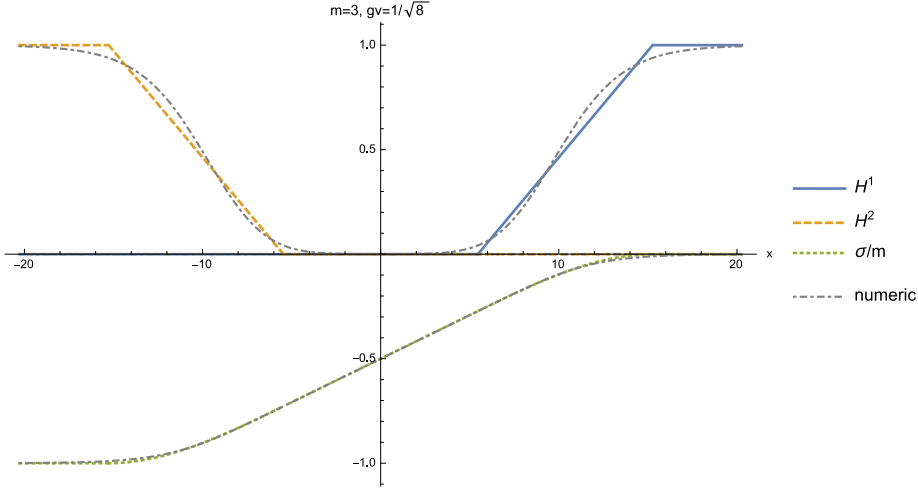}
\caption{\small Comparison of the crude model and numerical solution (gray dashed-dotted line). The parameters are fixed as $\sqrt{2}gv = 1/2$ and $m=3$, which corresponds to the values $R_c \approx 10.9$ and $R_s \approx 9.8$.
}
\label{fig:comparison}
\end{figure}

However, not all of the energy has been accounted for. Notice that at the core and in the vacua our fields are exact BPS solutions, while in the skin only the second BPS equation  \refer{eq:bpsw2} is satisfied. This means that besides the tension $T$ there is an additional energy contribution $E_{\small \mbox{extra}}$ coming from the violation of of the first BPS equation.
\begin{equation}
E_{\small \mbox{extra}} = \int\limits_{\small \mbox{skin}} \left| \partial_x H+\sigma H-HM\right|^2
= \frac{2v^2}{R_s}-\frac{4g^2v^4}{15}R_s+\frac{34 g^4v^6}{2835}R_s^3\,,
\end{equation}
where we have used the relation \refer{eq:relation}. We want this term to be as small as possible. Remarkably, $E_{\small \mbox{extra}}$ is strictly positive with a unique minimum at 
\begin{equation}
R_s = \frac{c_s}{g v}\,, \hspace{5mm} c_s = \sqrt{\frac{63+3\sqrt{2226}}{17}}\,.
\end{equation}
Thus, using our crude model we managed to find formulas relating width of the core $R_c$ and of the skin $R_s$ to the parameters of the model as 
\begin{equation}\label{eq:radii}
R_c= \frac{m}{g^2v^2}-\frac{4c_s}{3g v}\,, \hspace{5mm} R_s = \frac{c_s}{g v}\,, \hspace{5mm} c_s \approx 3.47\,,
\end{equation} 
which are in full accordance with the previous estimates found in \cite{Shifman, Eto1}. As per our intuitive understanding, the core develops only when tension of the domain wall is sufficiently big. More precisely, the crude model gives us an estimate $T_{\small \mbox{min}} \approx 4.63 g v^3$, which corresponds to $R_c = 0$ value.\footnote{One can imagine modified versions of the crude model, where the Higgs fields in the skin can deviate from a straight line. For example, if we assume sine-like profile and follow the same procedure as in the linear case, we obtain the formulas
\begin{equation}
R_c = \frac{m}{g^2v^2}-\frac{5}{4}R_s\,, \hspace{5mm} R_s = \frac{c_s}{g v}\,, \hspace{5mm} c_s \approx \pi \sqrt{\frac{2}{75\pi^2-705}\bigl(13+\sqrt{1200\pi^2-11111}\bigr)}\,,
\end{equation}
which gives slightly larger estimate $T_{\tiny \mbox{min}} \approx 5.9 g v^3 $. Intuitively, closer to the actual solution the profile in the skin is, more energy is shifted from the core to the skin, giving us higher values for $T_{\tiny \mbox{min}}$. In that sense, the value of the crude model $T_{\tiny \mbox{min}} \approx 4.63 g v^3$ can be seen as a lower bound.} The numerical analysis confirms that the formulas in Eq.~\refer{eq:radii} are working. For illustration, we show the comparison of the crude model with a particular numerical solution on Fig.~\ref{fig:comparison}.

\newpage

\section{Chains of walls}\label{sec:results}

Any domain wall solution of the master equation \refer{eq:masterw} is fully specified by the source term $\Omega_0$, which in turn depends on the mass matrix $M$ and on (absolute value of) the moduli matrix $H_0$. However, given $\Omega_0$ it is not possible to reconstruct $M$ and $H_0$ unambiguously, as we can always include any number of zero elements inside $H_0$ or degenerate entries in $M$. This effectively lifts the solution to a model with an arbitrarily large number of Higgs fields $N_F$.
Thus, in the rest of the paper, when we specify a mass matrix $M$ or a moduli matrix $H_0$ we always refer to the \emph{minimal} model, which can be ascertained unambiguously.

In this section, we present an infinite amount of exact solutions for the master equation \refer{eq:masterw}. We do this by first presenting two new single wall solutions -- or one-chains -- and then we show that simple addition of these yields new multi-wall solutions -- or multi-chains -- provided that certain restriction on parameters holds.

The existence of such solutions might be surprising, given the non-linearity of the master equation. However, it is important to keep in mind, that for each solution the source term $\Omega_0$ is different. To be precise, if we link a single wall solution to an existing chain, two things happen. First, the (minimal) model, where the solution lives, is enlarged. If we denote the mass matrix of the $n$-chain as $M^{(n)}$ and a mass matrix of the link as $M^{(1)}$, the new chain's mass matrix is given as $M^{(n+1)} = M^{(n)}\oplus M^{(1)}$, where by $\oplus$ we mean all possible pairings of elements from both matrices.\footnote{That is, for example
\begin{equation}
\mbox{diag}(m_1, m_2) \oplus \mbox{diag}(n_1,n_2 ) = \mbox{diag}(m_1+n_1, m_1+n_2, m_2+n_1, m_2+n_2)\,.
\end{equation}} The moduli matrices of the chain and newly added link combine in a much more complicated fashion.
 
Second, the constraint on the parameters, which is present to ensure the validity of the solution, becomes more strict. Roughly speaking, the maximum tension per individual wall is decreased. As we will argue, this tightening occurs primarily to prevent a violation of the no-go policy:
\begin{equation*}
\mbox{\bf No cores for exact walls}
\end{equation*}
This is the key observation of this paper. As far as we know there is no exact solution of domain wall with a core. If such solutions exist at all, they are certainly out of the reach of methods used in this paper. The  search for an exact domain wall solution with a core remains an interesting open problem.

Another peculiarity of exact solutions presented in this paper is that there are always \emph{non-elementary} walls. Expect few cases, which we show in the next subsection and which were already reported elsewhere \cite{Sakai3}, we have found no exact solution with isolated elementary walls. In fact, the number of \emph{separable} wall through the moduli parameters in the solutions themselves is always less than the number of elementary domain walls of the model. In short, our solutions cover only subspaces of the full moduli space.

In the following two subsections, we present all one-chains, two-chains and three-chains. We will shortly discuss their properties, but we mainly focus on three characteristics: 1) the minimal number of flavors, 2) the condition that ensures their validity and 3) check that they have no cores. Based on these particular examples, in the third subsection we provide general formulas of these characteristic for arbitrary $N$-chains.   

\subsection{Exact one-chains}

In the paper \cite{Sakai3} three exact single wall solutions were presented. In our notation, they read
\begin{align}
\label{eq:ex1} u_I^{2F}(x) & = 2\log\Bigl(1+e^{-m x}\Bigr)\,, & 
T  & = \sqrt{2}gv^3, \\ 
\label{eq:ex2} u_{II}^{2F}(x) & = 2\log\Bigl(1+\sqrt{6}e^{-mx}+e^{-2m x}\Bigr)\,, & 
T & = \sqrt{8}gv^3, \\ 
\label{eq:ex3} u_{III}^{2F}(x) & =  3\log\Bigl(1+e^{-m x}\Bigr)\,, & 
T & = \frac{3}{2}\sqrt{2}gv^3.
\end{align}
Here we have decorated each solution by the number of flavors in the superscript and by a roman numeral in the subscript for ease of reference.
We have also kept the parameter $m$, although $m = \sqrt{2}g v$ holds for all three solutions, to keep the formulas simple. 

Solutions \refer{eq:ex1}-\refer{eq:ex3} are remarkable for many reasons, two of which are particularly relevant here. First, they are the only instances of exact \emph{elementary} domain walls. As we stated, all other exact solutions presented in this paper are non-elementary walls.
Secondly,
all walls are core-less, since $T< T_{\small \mbox{min}} \approx 4.63 g v^3$. In other words, these walls are not heavy enough to have cores, as we discussed in subsection \ref{sec:crude}. 

Can we make chains out of $u_{I}^{2F}$, $u_{II}^{2F}$ or $u_{III}^{2F}$? In general, no.
The only exception is a remarkable double wall solution also reported in  \cite{Sakai3}, which in our notation reads
\begin{align}
u_I^{3F}(x) & = 2\log\bigl(1+\sqrt{6+e^{R}}\, e^{-m x}+e^{-2m x}\bigr)\,,  \\
\Omega_0 & = 1+e^{R}e^{-2m x}+ e^{-4m x}\,, \hspace{3mm} m = \sqrt{2}g v\,.
\end{align}
The parameter $R/m$ can be interpreted as a separation of walls if $R \gg m$, while for $R \sim m$ or less the walls are compressed.  As already noted in \cite{Sakai3}, $u_{I}^{3F}$ can be rewritten as:
\begin{align}
\label{eq:pseudolin2} u_I^{3F}(x) & = u_I^{2F}(x+S)+u_I^{2F}(x-S)\,, \hspace{3mm} S=\frac{1}{2m}\mbox{arcosh}\bigl(2+e^{R}/2\bigr)\,.
\end{align}
This is a first example of a two-chain. Indeed, $u_I^{3F}$ is a solution of a 3F model, written as a sum of two 2F domain walls and the mass matrix equals to the direct sum: 
$M = \mbox{diag}(0,m,2m) = \mbox{diag}(0,m)\oplus\mbox{diag}(0,m)$. A peculiar feature of $u_{I}^{3F}$ is that it is undefined for 
$\abs{S}< \abs{S_{\footnotesize \mbox{min}}} \equiv \tfrac{1}{2m}\mbox{arcosh}(2)$, which would cause $R$ to be complex.  
In fact, at $\abs{S}=\abs{S_{\footnotesize \mbox{min}}}$ the solution $u_{I}^{3F}$ becomes $u_{II}^{2F}$, which is an elementary wall. This peculiarity, however, seems to be isolated only to this 2-chain.

With the exception of  \refer{eq:pseudolin2}, no other chains can be constructed out of \refer{eq:ex1}-\refer{eq:ex3}.
Let us illustrate this on  $u_{III}^{2F}$.
Quick calculation reveals that the function $u_{III}^{2F}(x-S)+u_{III}^{2F}(x+S)$ is a solution to the master equation for the source term
\begin{equation}\label{eq:falsechain}
\Omega_0 = 1-9 e^{-2m x}+e^{R}e^{-3m x}-9e^{-4m x}+e^{-6m x}\,,
\end{equation}
where $R$ is again a function of $S$.
As we see, 
 $H_0$ is ill-defined given the negative signs in front of $e^{-2m x}$ and $e^{-4m x}$. Indeed, these would imply complex positions for first and third elementary domain walls (see Eq.~\refer{eq:exes}). Therefore, $u_{III}^{2F}$ cannot be chained. Same problems arise for any different combination of solutions \refer{eq:ex1}-\refer{eq:ex3}.

At this point we need new exact solutions. There are two, which we have found.
The first is obtained by relaxing the constraint on the parameter $m$ in $u_I^{2F}(x)$:
\begin{align}\label{eq:link1}
u_{II}^{3F}(x) & = 2\log\bigl(1+e^{-m x}\bigr)\,, \\
\Omega_0 & = 1+2e^{-m x}\bigl(1-\tfrac{m^2}{2g^2v^2}\bigr)+e^{-2mx}\,, \hspace{5mm} 
m \leq \sqrt{2}g v\,.
\end{align} 
Although very similar to $u_I^{2F}$, the mass parameter in $u_{II}^{3F}$ can be chosen arbitrarily within the range $0<m^2\leq 2g^2v^2$. This solution lives in 3F model with the mass matrix $M^{(1)} = -\mbox{diag}(0,m,2m)/2$. 
From Eq.~\refer{eq:exes}  we can calculate (naive) positions of elementary walls to be
\begin{equation}\label{eq:elw}
x_1 = \tfrac{1}{m}\log\Bigl(2- \tfrac{m^2}{g^2v^2}\Bigr)\,, \hspace{5mm} x_2 = -\tfrac{1}{m}\log\Bigl(2- \tfrac{m^2}{g^2v^2}\Bigr)\,.
\end{equation}
But since $x_1 < x_2$ both walls are compressed and there is, in fact, only a single wall with the tension $T_{II}^{3F} \equiv m v^2 \leq \sqrt{2}g v^3$ located at $m x_1 + m x_2 = 0$. In the limit $m^2 \to 2g^2v^2$ the compression becomes infinite $x_1 \to -\infty$, $x_2 \to \infty$ and we return to the original solution $u_I^{2F}$. 

Unlike $u_{I}^{3F}$, our new solution does not have a moduli, which controls separation of elementary walls.
In other words, this solution covers only a subspace of the full moduli space of the (minimal) model, where it lives. 
Also, notice that the amount of compression for both elementary walls is very specific $m$-dependent number. If the numbers $x_1$ and $x_2$ in Eq.~\refer{eq:elw} were only slightly different, no corresponding exact solution is known to the author. The reason -- if there is one -- why only such a special arrangement of elementary walls allows an exact solution, while other arrangements do not, remains elusive.

The second exact solution is constructed similarly by relaxing mass constraint in $u_{III}^{2F}$:
\begin{align}\label{eq:link2}
u_{I}^{4F}(x) & = 3\log\bigl(1+e^{-m x}\bigr)\,, \hspace{5mm} 
m \leq \sqrt{2}g v\,, \\
\Omega_0 & = 1+3e^{-m x}\bigl(1-\tfrac{m^2}{2g^2v^2}\bigr)+3\bigl(1-\tfrac{m^2}{2g^2v^2}\bigr)e^{-2mx} + e^{-3mx}\,.
\end{align} 
This solution lives in 4F model with the mass matrix $M^{(2)} = -\mbox{diag}(0,m,2m,3m)/2$. The ordering of elementary walls
\begin{equation}
x_1 = \tfrac{1}{m}\log\Bigl(3- \tfrac{3m^2}{2g^2v^2}\Bigr) < 0\,, 
\hspace{5mm} x_2 = 0 \,,
\hspace{5mm} x_3 = -\tfrac{1}{m}\log\Bigl(3- \tfrac{3m^2}{2g^2v^2}\Bigr) >0 \,,
\end{equation}
again implies that there is only one compressed wall at the origin. The original solution is restored in the limit $m^2 \to 2 g^2 v^2$, as before. The tension is never greater than $T_{I}^{4F} \equiv 3 m v^2/2 \leq 3\sqrt{2} g v^3/2$.

Both our new solutions $u_{II}^{3F}$ and $u_I^{4F}$ are obviously core-less, since their are no heavier than $u_{I}^{2F}$ and $u_{III}^{2F}$, respectively. But before we discuss two-chains made of $u_{II}^{3F}$ and $u_{I}^{4F}$, let us introduce a simple check that given exact solution has no core. The trick is to find the maximum value of the tension density 
\begin{equation}
{\mathcal T}_{\footnotesize \mbox{max}} \equiv \frac{v^2}{2}u^{\prime\prime}(0)\Big|_{\footnotesize \mbox{coincident walls}}
\end{equation} with all walls fixed at, say, the origin $x=0$. 
Since inside the core (as discussed in subsection \ref{sec:crude}) the tension density is practically equal to $g^2v^4$, it is sufficient to show that ${\mathcal T}_{\footnotesize \mbox{max}}$ is lower that this value.  
Indeed, short calculation reveals that all exact single wall solutions given in this subsection have their maximal values
\begin{gather}
{\mathcal T}_{I}^{2F}(0)  = 0.5 g^2v^4\,, \hspace{5mm}
{\mathcal T}_{II}^{2F}(0) \approx 0.89 g^2v^4\,, \hspace{5mm}
{\mathcal T}_{III}^{2F}(0) = 0.75 g^2v^4\,,  \\
{\mathcal T}_{II}^{3F}(0) = \frac{m^2 v^2}{4}\leq 0.5 g^2 v^4\,, \hspace{5mm}
{\mathcal T}_{I}^{4F}(0) = \frac{3m^2 v^2}{8}\leq 0.75 g^2 v^4
\end{gather}
markedly lower than $g^2v^4$.

\subsection{Exact two-chains and three-chains}

Let us first investigate  a two-chain made of equal tension $u_{II}^{3F}$ walls. That is:
\begin{align}\label{eq:5Fsol}
u_{I}^{5F}(x) & = u_{II}^{3F}(x+S;m)+u_{II}^{3F}(x-S;m)\,, \\ 
\Omega_0 &= 1+  4e^{-m x}\cosh(m S)\bigl(1-\tfrac{m^2}{2 g^2v^2}\bigr) +
2e^{-2m x}\bigl(2+\cosh(2mS)-\tfrac{2m^2}{g^2v^2}\bigr) \nonumber \\
\label{eq:5Fsol2} & + 4e^{-3m x}\cosh(m S)\bigl(1-\tfrac{m^2}{2 g^2v^2}\bigr) +e^{-4m x}\,.
\end{align}
This two-chain lives in 5F model with the mass matrix 
\begin{equation*}
M = M^{(1)}(m)\oplus M^{(1)}(m) = -\mbox{diag}(0,m,2m,3m,4m)/2\,.
\end{equation*}
Notice that the (minimal) number of flavors of $u_I^{5F}$ is not always five, but can be lower for special values of $m$ or $S$ (or both).
For example, taking the limit $m^2 \to 2g^2 v^2$ eliminates two factors from its  $\Omega_0$, which means that the resulting solution has only three flavors. Incidentally, this solution is $u_{I}^{3F}$. Also, by choosing $S=\tfrac{1}{2m}\arccosh\bigl(2-\tfrac{2m^2}{g^2v^2}\bigr)$ we can nullify third term in \refer{eq:5Fsol}, which gives us a new 4F solution
\begin{align}
u_{II}^{4F}(x) & = 2\log\Bigl(1+2e^{-m x}\sqrt{\tfrac{m^2}{g^2v^2}-\tfrac{1}{2}}+e^{-2mx}\Bigr)\,, 
\hspace{3mm} \frac{1}{4}\leq \frac{m^2}{2g^2v^2} \leq 1\,,\\
\Omega_0 & = 1 + 4e^{-mx}\Bigl(1+e^{-2mx}\Bigr)\bigl(1-\tfrac{m^2}{2g^2v^2}\bigr)\sqrt{\tfrac{m^2}{g^2v^2}-\tfrac{1}{2}}
+e^{-4m x}\,.
\end{align} and taking again the limit $m^2 \to 2g^2 v^2$ in this solution gives us $u_{II}^{2F}$. 

Generally, all chains can be reduced in a similar manner to a sequence of new exact solutions, which however possess fewer parameters than their parent solutions. We will investigate these `reductions' in detail in Sec.~\ref{sec:grinder}, since their exact structure becomes quickly very complicated for a higher number of flavors. 
In the remainder of this section, we  will skip these considerations entirely, for brevity.

Before going further, let us discuss a methodology, how we determine  under which  condition \emph{any} chain is a valid solution of the master equation \refer{eq:masterw}. For simple solutions such as $u_{I}^{5F}$ and $u_{II}^{4F}$ this can be done by inspection, but for more complicated ones, we need to establish a clear criterium. 

As we saw in the example of a false two-chain made out of $u_{III}^{2F}$ in Eq.~\refer{eq:falsechain}, the
issue is that we have to guarantee that moduli matrix $H_0$ is well defined or, more simply, that all coefficients multiplying exponential factors in $\Omega_0$  are positive numbers.
These coefficients, however, are generally complicated functions of distances between walls, which makes the analysis very cumbersome. Fortunately, in order to obtain a general condition for any arrangement of walls, we only need to look at the coincident case, where all walls sit on the same point on the $x$-axis, say the origin. At this limit, all coefficients in $\Omega_0$ reach their lowest values.

Why it is so? Let us illustrate this on an example of a generic two-chain made of single-wall solutions $u_1$ and $u_2$, where
\begin{equation}
\frac{1}{2g^2v^2}\partial_x^2 u_1 = 1-\Omega_{0}^{(1)}e^{-u_1}\,, \hspace{5mm}
\frac{1}{2g^2v^2}\partial_x^2 u_2 = 1-\Omega_{0}^{(2)}e^{-u_2}\,.
\end{equation} 
The master equation for $u_1+u_2$ can be rewritten as
\begin{align}
\frac{1}{2g^2v^2}\partial_{x}^2(u_1+u_2) & = 2-\Omega_{0}^{(1)}e^{-u_1}-\Omega_{0}^{(2)}e^{-u_2} \nonumber \\
& = 
1- \Omega_{0}^{(1)}\Omega_{0}^{(2)}e^{-u_1-u_2}+\bigl(1-\Omega_{0}^{(1)}e^{-u_1}\bigr)
\bigl(1-\Omega_{0}^{(2)}e^{-u_2}\bigr) \nonumber \\
& = 1- \Omega_{0}^{(1)}\Omega_{0}^{(2)}e^{-u_1-u_2} +\frac{1}{4g^4v^4}\partial_{x}^2u_1\partial_{x}^2u_2 \nonumber \\
& = 1- \Bigl( \Omega_{0}^{(1)}\Omega_{0}^{(2)}-\frac{1}{4g^4v^4}\partial_{x}^2u_1\partial_{x}^2u_2e^{u_1+u_2}\Bigr)e^{-u_1-u_2}\,.
\end{align}
Thus, the two-chain  $u_1+u_2$ solves the master equation with the source term 
\begin{equation}
\Omega_0^{(1+2)} = \Omega_{0}^{(1)}\Omega_{0}^{(2)}-\frac{1}{4g^4v^4}\partial_{x}^2u_1\partial_{x}^2u_2e^{u_1+u_2}\,.
\end{equation}
$\Omega_0^{(1+2)}$ has two parts. The first part is manifestly positive since  $\Omega_{0}^{(1)}\Omega_{0}^{(2)}$ has only positive coefficients.
The second part is proportional to tension densities of both solutions and hence (given the minus sign) is manifestly negative. Thus, the coefficients in $\Omega^{(1+2)}$ are lowest if the second term is (in an absolute sense) largest. But this happens precisely when domain walls from first and second solution are coincident because the product of their tension densities $\sim \partial_x^2u_1\partial_x^2u_2$ is largest there.
Hence, our claim holds. Notice that this argument can be trivially extended to any number of walls.
Therefore, the criterium, which ensures well-defined $H_0$ for any exact solution, is established by looking at the lowest coefficient in $\Omega_0$ at the coincident point. 

In case of $u_{I}^{5F}$ the coincident limit reads
\begin{equation}
\Omega_0 \xrightarrow[]{S\to 0} 1+ 4\bigl(1-\tfrac{m^2}{2g^2v^2}\bigr)e^{-mx}+2\bigl(3-\tfrac{2m^2}{g^2v^2}\bigr)e^{-2mx}
+4\bigl(1-\tfrac{m^2}{2g^2v^2}\bigr)e^{-3mx}+e^{-4mx}\,,
\end{equation}
from which we see\footnote{But see the discussion bellow Eq.~\refer{eq:5Fsol2}.}
\begin{equation}\label{eq:5Fmcond}
m^2 \leq 3g^2v^2/2\,.
\end{equation}
Notice that this is more stringent than for individual one-chains, where the condition was $m^2 \leq 2 g^2 v^2$. This is intuitively understandable from the point of view of no-cores-for-exact-solutions philosophy. Since the tensions of individual links in the chain add up linearly, the restriction on them must be larger, compared to restrictions on links themselves. If that were not the case, a sufficiently long chain would eventually break the threshold for developing a core.
And indeed, with the more stringent restriction the maximum tension density for our 2-chain is well bellow the threshold
\begin{equation}\label{eq:5FT}
{\mathcal T}_{I}^{5F}(0) = \frac{v^2 m^2}{2} \leq \frac{3 g^2 v^4}{4}\,.
\end{equation}

Let us now consider a two-chain made of two $u_{II}^{3F}$ walls with unequal tensions:
\begin{equation}
u_{IV}^{9F}(x) = u_{II}^{3F}(x-S_1;m_1)+u_{II}^{3F}(x-S_2;m_2)\,.
\end{equation}
This two-chain is a special case of a more general solution designated as $u_{IV}^{9F}(x)$, which is studied in detail in Sec~\ref{sec:grinder}.
We count 9 flavors. The mass matrix
\begin{align}
M & = M^{(1)}(m_1) \oplus M^{(1)}(m_2) \nonumber \\ & = -\mbox{diag}(0, m_1, m_2, m_1+m_2, 2m_1, 2m_2, 2m_1+m_2, m_1+2m_2, 2m_1+2m_2)/2 \nonumber
\end{align}
is again given by considering all possible pairings between matrix elements of its single wall constituents. 
At the
coincident point ($S_1=S_2 =0$) we have
\begin{align}
\Omega_0 & = 1+2\bigl(1-\tfrac{m_1^2}{2g^2v^2}\bigr)\bigl(1+e^{-2m_2 x}\bigr)e^{-m_1 x}+2\bigl(1-\tfrac{m_2^2}{2g^2v^2}\bigr)\bigl(1+e^{-2m_1 x}\bigr)e^{-m_2 x}
\nonumber \\
& +e^{-2m_1 x}+e^{-2m_2 x}+4\Bigl(1-\tfrac{m_1^2+m_2^2}{2g^2v^2}\Bigr)e^{-(m_1+m_2) x}+e^{-2(m_1+m_2)x}\,.
\end{align}
Thus, the solution $u_{IV}^{9F}$ is valid if
\begin{equation}
m_1^2+m_2^2 \leq 2 g^2 v^2\,,
\end{equation}
which is again more severe than the equivalent for one-chains.
The limit on the tension density is
\begin{equation}
{\mathcal T}_{IV}^{9F}(0) = v^2\frac{m_1^2+m_2^2}{4} \leq 0.5 g^2 v^4\,.
\end{equation} 
Notice that taking limit $m_1 = m_2 \equiv m $ does neither reproduce the condition \refer{eq:5Fmcond} nor \refer{eq:5FT}. This is why we have dealt with both cases separately. In fact, many characteristics of the chain change discontinuously, for different levels of mass-degeneracies. For example, the number of elementary walls is now eight, indicating that both walls are composites of four, rather than two elementary walls, as is the case when we look at each $u_{II}^{3F}$ link independently. Thus, when $m_1\not = m_2$ the act of combining both solutions changes their nature. Nothing of this sort happens in equal tension case.

Instead of continuing the discussion for 2-chains involving combinations of  $u_{II}^{3F}$ and $u_{I}^{4F}$, we rather collect all the relevant information about them into the table \ref{tab:twochains}. There we show number of flavors $N_F$, condition on parameter(s) and maximum value of the tension density ${\mathcal T}_{\footnotesize \mbox{max}}/g^2v^4$ for each solution.

\begin{table}[!h]
\begin{tabular}{c|c|c|c}
\hline \hline
chain & $N_F$ &  cond. & ${\mathcal T}_{\footnotesize \mbox{max}}/g^2v^4$  \\
\hline \hline
$\mathbf{1}(m)$ & 3F &  $m^2 \leq 2 g^2 v^2$ & $1/2$ \\
\hline
$\mathbf{2}(m)$ & 4F &  $m^2 \leq 2 g^2 v^2$ & $3/4$ \\
\hline \hline
$\mathbf{1}(m)\oplus \mathbf{1}(m)$ & 5F &  $m^2 \leq 3 g^2 v^2/2$ & $3/4$ \\
\hline
$\mathbf{1}(m)\oplus \mathbf{2}(m)$ & 6F &  $m^2 \leq 4 g^2 v^2/3$ & $5/6$ \\
\hline
$\mathbf{2}(m)\oplus \mathbf{2}(m)$ & 7F &  $m^2 \leq 10 g^2 v^2/9$ & $5/6$ \\
\hline
$\mathbf{1}(m_1)\oplus \mathbf{1}(m_2)$ & 9F & $m_1^2+m_2^2 \leq 2 g^2 v^2$ & $1/2$ \\
\hline
$\mathbf{1}(m_1)\oplus \mathbf{2}(m_2)$ & 12F &  $m_1^2+m_2^2 \leq 2 g^2 v^2$ & $3/4$ \\
\hline
 $\mathbf{2}(m_1)\oplus \mathbf{2}(m_2)$ & 16F & $m_1^2+m_2^2 \leq 2 g^2 v^2$ & $3/4$ \\
\hline \hline
\end{tabular}
\caption{\small All 1-chains and 2-chains.}
\label{tab:twochains}
\end{table}

In Tab.~\ref{tab:twochains} we have introduced a compact notation $\mathbf{1}(m)\equiv u_{II}^{3F}(x;m)$ and $\mathbf{2}(m)\equiv u_{I}^{4F}(x;m)$ for the single wall solutions. To indicate a two-chain we use $\oplus$ symbol, which represents a plain sum for solutions, but direct sum for mass matrices $M^{(1)}(m)$ and  $M^{(2)}(m)$. We do not show $\Omega_0$, since the formulas are too long. However, one can easily reconstruct both the solution itself and the corresponding source term from the prescription given in the first column in the table and from the identity
\begin{equation}
\Omega_0 = e^{u(x)}\Bigl(1-\frac{1}{2g^2v^2}\partial_{x}^2 u(x)\Bigr)\,,
\end{equation} 
which holds for any solution $u(x)$ of the master equation.

We list the same results for the 3-chains in table \ref{tab:threechains}.
 
\begin{table}[!h]
\begin{tabular}{c|c|c|c}
\hline \hline
chain & $N_F$ &  cond. & ${\mathcal T}_{\footnotesize \mbox{max}}/g^2v^4$  \\
\hline \hline
$\mathbf{1}(m)\oplus\mathbf{1}(m)\oplus\mathbf{1}(m)$ & 7F &  $m^2 \leq 10 g^2 v^2/9$ & $5/6$ \\
\hline
$\mathbf{1}(m)\oplus\mathbf{1}(m)\oplus\mathbf{2}(m)$ & 8F &  $m^2 \leq g^2 v^2$ & $7/8$ \\
\hline
$\mathbf{1}(m)\oplus\mathbf{2}(m)\oplus\mathbf{2}(m)$ & 9F &  $m^2 \leq 7 g^2 v^2/8$ & $7/8$ \\
\hline
$\mathbf{2}(m)\oplus\mathbf{2}(m)\oplus\mathbf{2}(m)$ & 10F &  $m^2 \leq 4 g^2 v^2/5$ & $9/10$ \\
\hline
$\mathbf{1}(m_1)\oplus\mathbf{1}(m_1)\oplus\mathbf{1}(m_2)$ & 15F &  $4m_1^2+3m_2^2 \leq 6 g^2 v^2$ & $3/4$ \\
\hline
$\mathbf{1}(m_1)\oplus\mathbf{1}(m_2)\oplus\mathbf{2}(m_1)$ & 18F &  $3m_1^2+2m_2^2 \leq 4 g^2 v^2$ & $5/6$ \\
\hline
 $\mathbf{1}(m_1)\oplus\mathbf{1}(m_1)\oplus\mathbf{2}(m_2)$ & 20F & $4m_1^2+3m_2^2 \leq 6 g^2 v^2$ & $3/4$ \\
 \hline
 $\mathbf{1}(m_1)\oplus\mathbf{2}(m_2)\oplus\mathbf{2}(m_2)$ & 21F & $5m_1^2+9m_2^2 \leq 10 g^2 v^2$ & $5/6$ \\
  \hline
 $\mathbf{1}(m_1)\oplus\mathbf{2}(m_1)\oplus\mathbf{2}(m_2)$ & 24F & $3m_1^2+2m_2^2 \leq 4 g^2 v^2$ & $5/6$ \\
 \hline
$\mathbf{1}(m_1)\oplus\mathbf{1}(m_2)\oplus\mathbf{1}(m_3)$ & 27F & $m_1^2+m_2^2+m_3^2 \leq 2 g^2 v^2$ & $1/2$ \\
\hline
$\mathbf{2}(m_1)\oplus\mathbf{2}(m_1)\oplus\mathbf{2}(m_2)$ & 28F & $9m_1^2+5m_2^2 \leq 10 g^2 v^2$ & $5/6$ \\
\hline
$\mathbf{1}(m_1)\oplus\mathbf{1}(m_2)\oplus\mathbf{2}(m_3)$ & 36F & $m_1^2+m_2^2+m_2^3 \leq 2 g^2 v^2$ & $3/4$ \\
\hline
$\mathbf{1}(m_1)\oplus\mathbf{2}(m_2)\oplus\mathbf{2}(m_3)$ & 48F & $m_1^2+m_2^2+m_2^3 \leq 2 g^2 v^2$ & $3/4$ \\
\hline
$\mathbf{2}(m_1)\oplus\mathbf{2}(m_2)\oplus\mathbf{2}(m_3)$ & 64F & $m_1^2+m_2^2+m_2^3 \leq 2 g^2 v^2$ & $3/4$ \\
\hline \hline
\end{tabular}
\caption{\small All 3-chains.}
\label{tab:threechains}
\end{table}

\subsection{Exact $N$-chain}

Let us define an arbitrary $N$-chain as a sum of $N$ single wall solutions formally denoted as
\begin{equation}
u[\mathbf{X}] = \bigoplus_{i=1}^{N}\mathbf{X}_i(m_i)\,,
\end{equation}
where $\mathbf{X} \in \{\mathbf{1}\equiv u_{II}^{3F}, \mathbf{2}\equiv u_{I}^{4F}\}$. 
The parameters $m_i$ can be either same or different for each link ${\mathbf{X}}_i(m_i)$. But, as was the case for 2-chains and 3-chains, all characteristics of interest, such as a number of flavors $N_F$, condition ensuring the validity of a solution and ${\mathcal T}_{\footnotesize \mbox{max}}$, change discontinuously across degeneracies. This means that taking limits of parameters $m_i$ in formulas found for the non-degenerate case does not give correct formulas for partially degenerate cases.  Therefore, the goal of this subsection is to develop general formulas for all three observables, which apply for all possible degeneracies of parameters $m_i$ from the beginning.

Let us start with the number of flavors $N_F$. The rule of thumb turns out to be that for the links with the same $m_i$, the number of flavors \emph{minus one} add up, while for links with different $m_i$, flavors multiply. Let us illustrate this on two extreme examples. First, consider a  totally degenerate chain made out of  $N-l$ $\mathbf{1}$-links and $l$ $\mathbf{2}$-links:
\begin{equation}
\Bigl(\bigoplus_{i=1}^{N-l}\mathbf{1}(m)\Bigr)\oplus \Bigl(\bigoplus_{j=1}^{l}\mathbf{2}(m)\Bigr) \equiv \mathbf{1}^{N-l}(m)\oplus \mathbf{2}^{l}(m)\,.
\end{equation}
Here we introduced a shorthand notation $\mathbf{X}^{n}(m) \equiv \bigoplus_{i=1}^{n}\mathbf{X}(m)$. The solution at the coincident point is given as 
\begin{equation}
 u=  \log\Bigl(1+e^{-m x}\Bigr)^{2N+l}\,.
\end{equation}
The number of flavors $N_F$ is simply the number of distinct exponentials in $\Omega_0$, or equivalently in $e^{u}$.
This gives us the result 
\begin{equation}\label{eq:flinear}
N_F = 2N+l+1\,,
\end{equation}
which is simply the number of different powers of $e^{-mx}$ in $(1+e^{-mx})^{2N+l}$, including zero power.
Notice that the number of elementary domain walls can be rewritten as $N_F -1= 2(N-l)+3 l$. At the same time, $\mathbf{1}$-link is composed of two, while $\mathbf{2}$-link is composed of three elementary domain walls. Therefore, in totally degenerate case the elementary walls which simply add up. 

For a second example let us consider a totally non-degenerate chain:
\begin{equation}
\Bigl(\bigoplus_{i=1}^{N-l}\mathbf{1}(m_i)\Bigr)\oplus \Bigl(\bigoplus_{j=1}^{l}\mathbf{2}(m_j)\Bigr)  \Rightarrow u = \log \biggl(\prod_{i=1}^{N-l} \bigl(1+e^{-m_i x}\bigr)^2\prod_{j=1}^{l}\bigl(1+e^{-m_j x}\bigr)^3\biggr)\,.
\end{equation}
Counting the number of distinct exponential in $e^{u}$ we obtain
\begin{equation}\label{eq:fmult}
N_F =  3^{N-l}4^{l}
\end{equation}
Here we see the multiplication rule: Each  $\mathbf{1}$ links contributes three flavors and each $\mathbf{2}$ link gives four flavors.

A generic case can be characterized by partial degeneracy of the parameters $m_i$ into $k\leq N$ groups of size $d_i$, where $N = \sum_{i=1}^{k}d_i$. Within each group $d_i -l_i$ are $\mathbf{1}$-links and $l_i$ are $\mathbf{2}$-links. The solution at the coincident point of such chain reads
\begin{equation}\label{eq:genericN}
\bigoplus_{i=1}^{k}\bigl(\mathbf{1}^{d_i-l_i}(m_i)\oplus \mathbf{2}^{l_i}(m_i)\Bigr) \Rightarrow
u = \log\biggl(\prod\limits_{i=1}^{k}\Bigl(1+e^{-m_i x}\Bigr)^{2d_i+l_i}\biggr)\,.
\end{equation}
Within a degenerate group the number of flavors follow the additive rule of Eq.~\refer{eq:flinear}, that is we have $2d_i+l_i+1$ distinct exponentials for each group. These numbers must are then multiplied as per the rule in Eq.~\refer{eq:fmult}. Thus, the most general formula for arbitrary $N$-chain reads
\begin{equation}\label{eq:Nflavors}
\boxed{
N_F = \prod\limits_{i=1}^{k}\bigl(2d_i+l_i+1\bigr)\,.
}
\end{equation}
It is easy to check that the results in the second columns in tables \ref{tab:twochains} and \ref{tab:threechains} are concurrent to the above formula.
 
Let us now establish the condition under which arbitrary $N$-chain is a valid solution of the master equation \refer{eq:masterw}. Following the discussion in the previous subsection, we must investigate the lowest coefficient in $\Omega_0$
at the coincident point. For the most general $N$ chain given in Eq.~\refer{eq:genericN} we have
\begin{equation}
\Omega_0 = \prod\limits_{j=1}^{k}\Bigl(1+e^{-m_j x}\Bigr)^{2d_j+l_j}\biggl(1 - \sum\limits_{i=1}^{k}\frac{(2d_i+l_i)m_i^2}{2g^2v^2}\frac{e^{-m_i x}}{\bigl(1+e^{-m_i x}\bigr)^2}\biggr)\,.
\end{equation} 
Now we extract the lowest coefficient. Let us first write down the largest factor coming from the negative half (the second term in the large parenthesis) of $\Omega_0$:
\begin{equation}
-\sum\limits_{i=1}^{k}\binom{2d_i+l_i-2}{d_i+\lfloor\tfrac{l_i}{2}\rfloor-1}\prod\limits_{j\not = i}\binom{2d_j+l_j}{d_j+\lfloor\tfrac{l_j}{2}\rfloor}\frac{(2d_i+l_i)m_i^2}{2g^2v^2}
\exp\bigl(-\sum\limits_{n=1}^{k}\bigl(d_n+\lfloor\tfrac{l_n}{2}\rfloor\bigr)m_n x\bigr)\,,
\end{equation}
where $\lfloor x\rfloor$ is the floor function. Simply put, we have expanded each bracket in the product and picked out the largest factor, which is in the middle of the binomial series.
The factor with the same power in the exponential in the positive half reads
\begin{equation}
\prod\limits_{i}^{k}\binom{2d_i+l_i}{d_i+\lfloor\tfrac{l_i}{2}\rfloor}\exp\bigl(-\sum\limits_{n=1}^{k}\bigl(d_n+\lfloor\tfrac{l_n}{2}\rfloor\bigr)m_n x\bigr)\,.
\end{equation}
Combining both factors together and demanding that the overall coefficient is non-negative we find the condition for arbitrary $N$-chain to be
\begin{equation}\label{eq:fullcond}
\boxed{
\sum\limits_{i=1}^{k}\frac{\bigl(d_i+\lfloor\tfrac{l_i+1}{2}\rfloor\bigr)\bigl(d_i+\lfloor\tfrac{l_i}{2}\rfloor\bigr)}{2d_i+l_i-1}m_i^2 \leq 2g^2v^2\,.}
\end{equation}

Let us consider few examples. The totally degenerate case is given by setting $k=1$, $m_1 \equiv m$ and $d_1 = N$. Taking $l_1 \equiv l$ out of $N$ walls to be $\mathbf{2}$-links we have the conditions
\begin{equation}
(l \mbox{ is even}) \hspace{3mm} m^2 \leq \frac{4 g^2v^2}{N+\tfrac{l}{2}}\Bigl(1-\frac{1}{2N+l}\Bigr)\,, \hspace{5mm} (l \mbox{ is odd}) \hspace{3mm} m^2 \leq \frac{4g^2v^2}{N+\tfrac{l+1}{2}}\,.
\end{equation}  
On the other hand in the fully non-degenerate case $k=N$, $d_i = 1$, $l_i \in \{0,1\}$ the general condition \refer{eq:fullcond} reduces to
\begin{equation}\label{eq:nondegcond}
\sum\limits_{i=1}^{N}m_i^2 \leq 2g^2v^2\,,
\end{equation}
which is curiously independent on all  $l_i$'s. Again, the validity of the condition \refer{eq:fullcond} for two-chains and three-chains can be checked by looking at the result in Tabs.~\ref{tab:twochains}-\ref{tab:threechains}.

The remaining task is to find a general formula for ${\mathcal T}_{\footnotesize \mbox{max}}$, defined as
\begin{equation}
{\mathcal T}_{\footnotesize \mbox{max}} \equiv \frac{v^2}{2}u^{\prime\prime}(0)\Big|_{\footnotesize \mbox{coincident walls}}
\end{equation}
For the generic $N$-chain of Eq.~\refer{eq:genericN} we find
\begin{equation}\label{eq:sum}
{\mathcal T}_{\footnotesize \mbox{max}} = \frac{v^2}{8}\sum\limits_{i=1}^{k}\bigl(2d_i+l_i\bigr)m_i^2\,.
\end{equation}
We need to find the maximum of this. From the formula \refer{eq:fullcond} we see that the upper bound for a $i$-th factor in the above sum is given as
\begin{equation}
(2d_i+l_i)m_i^2 \leq  2g^2v^2 \frac{(2d_i+l_i-1)(2d_i+l_i)}{\bigl(d_i+\lfloor\tfrac{l_i+1}{2}\rfloor\bigr)\bigl(d_i+\lfloor\tfrac{l_i}{2}\rfloor\bigr)}\,.
\end{equation}
This can be simplified to
\begin{equation}
(2d_i+l_i)m_i^2 \leq 8 g^2 v^2 \Bigl(1-\frac{1}{2\bigl(d_i+\lfloor\tfrac{l_i+1}{2}\rfloor\bigr)}\Bigr)\,.
\end{equation}
Obviously, the maximum of the sum of these numbers amounts to picking up such $i \equiv n$ for which $2d_n+l_n \geq 2d_i+l_i$ for $i=1,\ldots, k$. Then, the sum in Eq.~\refer{eq:sum} is maximized by taking $m_i = 0$ for $i\not = n$. In summary, the formula for maximum of the tension density for generic $N$-chain reads
\begin{equation}\boxed{
{\mathcal T}_{\footnotesize \mbox{max}} = g^2v^4 \Bigl(1-\frac{1}{2\bigl(d_n+\lfloor\tfrac{l_n+1}{2}\rfloor\bigr)}\Bigr)\,.}
\end{equation}
Its validity for 2-chains and 3-chains can be checked by comparing with the findings in the third columns of Tabs.~\ref{tab:twochains}-\ref{tab:threechains}. We see that ${\mathcal T}_{\footnotesize \mbox{max}}$ is always less than $g^2v^4$ as expected. The closest approach is achieved by fully degenerate $\mathbf{2}$ chains, that is $d_n = N$ and $l=N$:
\begin{align}
{\mathcal T}_{\footnotesize \mbox{max}} & = g^2v^4 \Bigl(1-\frac{1}{3N}\Bigr)\,, & (N \mbox{ is even}) \\
{\mathcal T}_{\footnotesize \mbox{max}} & = g^2v^4 \Bigl(1-\frac{1}{3N+1}\Bigr)\,. & (N \mbox{ is odd})
\end{align}
In other words $\abs{{\mathcal T}_{\footnotesize \mbox{max}}-g^2v^4} \sim 1/N$ for $N\gg1$. This might seem that for sufficiently big chains, the core should appear as the gap between ${\mathcal T}_{\footnotesize \mbox{max}}$ and $g^2v^4$ closes. However, this is misleading, because we are only looking at the height of the peak. But its full shape for fully degenerate $\mathbf{2}$-chain is
\begin{equation}
{\mathcal T} = \frac{v^2}{2}\partial_x^2 u  = \frac{m^2 v^2}{8}\frac{3N}{\cosh^2 (m x/2)} \leq g^2v^4 \frac{1}{\cosh^2 \bigl(gv\sqrt{\tfrac{2}{3N}}x\bigr)}\,, \hspace{3mm} N \gg 1\,.
\end{equation}
In other words, increase in $N$ causes only widening of the peak and not development of the plateau as seen in Fig.~\ref{fig:tensions}. This clearly confirms no-core rule for exact solution, as far as chains are concerned.

Let us close this section by commenting on the issue of infinite chains. Periodic domain wall solutions or other exact infinite wall configurations are potentially very interesting objects, worth studying on their own. However, the chains presented here cannot be extended to infinite case. The obstacle can be clearly seen from the condition \refer{eq:fullcond}, which implies that as $N\to \infty$, the parameters vanish $m_i \to 0$. More simply, if we look again at fully degenerate $\mathbf{2}$ chain as an illustrative example, we see that the amount of tension per wall decreases: 
\begin{equation}
T_{\footnotesize \mbox{max}}/N \sim \sqrt{\frac{6}{N}}\, g v^3\,, \hspace{5mm} N \gg 1\,.  
\end{equation}
In other words, taking $N$ to infinity, the chain disappears. 

\clearpage

\section{Hierarchy of exact solutions}\label{sec:grinder}

Rather than focusing on the particular type of exact solutions -- like chains of the previous section -- in this part we want to \emph{exhaust} all exact solutions, which can be written as the logarithm of a sum of exponentials. To that goal, in the first subsection, we present a convenient ansatz, or rather a class of ansatzes. We devote later subsections to investigate first few smallest domain wall configurations as examples, highlighting the main point of this section. That it, that all exact solutions form hierarchy, where any particular solution might be seen as a special limit in parameter space of another solution, living in the larger model. Unlike for chains, these relations turn out to be increasingly complicated as the minimal number of flavors of the underlying model increases. In particular, in this section we explore these relations in detail up to $N_F = 10$ flavors.  

For simplicity, throughout this section
we fix the effective gauge coupling parameter $\tilde g\equiv \sqrt{2}gv$ appearing in the master equation \refer{eq:masterw3} to  $\tilde g = 1$ unless otherwise stated. Any solution $u(x;m_i)$ at unit effective gauge coupling can be rescaled to arbitrary case as $u\bigl(\tilde g\, x;m_i/\tilde g\bigr)$.

\subsection{Ansatz}

Let us consider the function
\begin{equation}\label{eq:exsol}
u_1^{(N)}(x) \equiv 2\log F_N(x) \equiv 2\log\Bigl(1+\sum\limits_{i=1}^{N}\prod\limits_{j=1}^{i}e^{R_i-m
_i x}\Bigr)
\end{equation}
That is in particular
{\small
\begin{align}
\label{eq:exsol1} u_1^{(1)}(x) &= 2\log\Bigl(1+e^{R_1}e^{-m_1 x}\Bigr)\,, \\
\label{eq:exsol2} u_1^{(2)}(x) &= 2\log\Bigl(1+e^{R_1}e^{-m_1 x}+e^{R_1+R_2}e^{-(m_1+m_2)x}\Bigr)\,, \\
\label{eq:exsol3} u_1^{(3)}(x) &= 2\log\Bigl(1+e^{R_1}e^{-m_1 x}+e^{R_1+R_2}e^{-(m_1+m_2)x}+
e^{R_1+R_2+R_3}e^{-(m_1+m_2+m_3)x}\Bigr)\,.
\end{align}}
The reason why we choose the ansatz \refer{eq:exsol} in this way is that vanishing of any of $m_i$'s reduce the $N$-th solution to the $N-1$ case. Therefore, it is sufficient just to study strictly positive values of $m_i$'s. Also, all powers of $e^{-x}$ are automatically ordered since $0< m_1 < m_1+m_2< \ldots$. This allow us to easily estimate the positions of all (separable) walls, using similar arguments to those in Sec.~\ref{sec:model}: we look at points where a pair of neighbouring exponentials equals:
\begin{equation}\label{eq:exes2}
\tilde x_1 = \frac{R_1}{m_1}\,, \hspace{5mm} \tilde x_2 = \frac{R_2}{m_2}\,, \hspace{5mm} \ldots \hspace{5mm} \tilde x_N = \frac{R_N}{m_N}\,.
\end{equation} 
Of course, these numbers represents actual locations of domain walls if they are ordered $\tilde x_1 \gg \tilde x_2 \gg \ldots \gg \tilde x_N$, in the same way as we discussed for positions of elementary walls in Eq.~\refer{eq:exes}. 
Also notice that parameters $m_i$ entering \refer{eq:exsol} are not equal to the masses of the model, but they are related to them (see Eq.~\refer{eq:massmatrix}).

The crucial detail in our ansatz is the factor $2$ in front of the logarithm since it makes $\Omega_0$ automatically a finite sum of exponentials. Indeed, one can easily see that the general form is given as
\begin{equation}\label{eq:omega0}
\Omega_0^{(N)} = F_N^2 +2 (F_N^{\prime})^2-2 F_N^{\prime\prime} F_N\,.
\end{equation}
Let as expand first few lowest cases 
{\small \begin{align}
\Omega_0^{(1)} &= 1+2e^{R_1}e^{-m_1x}\bigl(1-m_1^2\bigr)+e^{2R_1}e^{-2m_1x}\,, \\
\Omega_0^{(2)} &= 1+2e^{R_1}e^{-m_1x}\bigl(1-m_1^2\bigr)+e^{2R_1}e^{-2m_1x}+2e^{R_1+R_2}e^{-(m_1+m_2)x}\bigl(1-(m_1+m_2)^2\bigr)\nonumber \\&+2e^{2R_1+R_2}e^{-(2m_1+m_2)}\bigl(1-m_2^2\bigr)+e^{2R_1+2R_2}e^{-(2m_1+2m_2)x}\,, \\
\Omega_0^{(3)} &= 1+2e^{R_1}e^{-m_1x}\bigl(1-m_1^2\bigr)+e^{2R_1}e^{-2m_1x}+2e^{R_1+R_2}e^{-(m_1+m_2)x}\bigl(1-(m_1+m_2)^2\bigr)\nonumber \\&+2e^{2R_1+R_2}e^{-(2m_1+m_2)}\bigl(1-m_2^2\bigr)+2e^{R_1+R_2+R_3}e^{-(m_1+m_2+m_3)x}\bigl(1-(m_1+m_2+m_3)^2\bigr)\nonumber \\ 
&+e^{2R_1+2R_2}e^{-(2m_1+2m_2)x}+2e^{2R_1+R_2+R_3}e^{-(2m_1+m_2+m_3)x}\bigl(1-(m_2+m_3)^2\bigr) \nonumber \\
&+2e^{2R_1+2R_2+R_3}e^{-(2m_1+2m_2+m3)x}\bigl(1-m_3^2\bigr) +
e^{2R_1+2R_2+2R_3}e^{-2(m_1+m_2+m_3)x}\,.
\end{align}}
We count 3, 6 and 10 independent flavors (terms) respectively, assuming generic $m_i$'s.
The (unordered) mass matrix and the real part of the moduli matrix are given in each case as:
\begin{align}
M^{(1)}  = -\mbox{diag}\Bigl(& 0, m_1, 2m_1\Bigr)/2\,, \\
M^{(2)}  = -\mbox{diag}\Bigl(& 0, m_1, 2m_1, m_1+m_2, 2m_1+m_2, 2m_1+2m_2\Bigr)/2\,, \\
M^{(3)}  = -\mbox{diag}\Bigl(& 0, m_1, 2m_1, m_1+m_2, 2m_1+m_2, 2m_1+2m_2,m_1+m_2+m_3, \nonumber \\
& 2m_1+m_2+m_3, 2m_1+2m_2+m_3,2m_1+2m_2+2m_3\Bigr)/2\,,
\end{align}
and 
{\small \begin{align}
H_0^{(1)}  = \Bigl(& 1, e^{R_1/2}\sqrt{2-2m_1^2}, e^{R_1}\Bigr)\,, \\
H_0^{(2)}  = \Bigl(& 1, e^{R_1/2}\sqrt{2-2m_1^2}, e^{R_1}, e^{(R_1+R_2)/2}\sqrt{2-2(m_1+m_2)^2},
e^{R_1+R_2/2}\sqrt{2-2m_2^2}, \nonumber \\ & e^{R_1+R_2}\Bigr)\,, \\
H_0^{(3)}=  \Bigl(& 1, e^{R_1/2}\sqrt{2-2m_1^2}, e^{R_1}, e^{(R_1+R_2)/2}\sqrt{2-2(m_1+m_2)^2},  \nonumber \\
& e^{R_1+R_2/2}\sqrt{2-2m_2^2}, e^{R_1+R_2}, e^{(R_1+R_2+R_3)/2}\sqrt{2-2(m_1+m_2+m_3)^2}, \nonumber \\
& e^{R_1+R_2/2+R_3/2}\sqrt{2-2(m_2+m_3)^2}, e^{R_1+R_2+R_3/2}\sqrt{2-2m_3^2}, e^{R_1+R_2+R_3}\Bigr)\,.
\end{align}}
Inspection of the above moduli matrices reveals that in order to have well defined solution the parameters $m_1$, $m_2$ and $m_3$ must be constrained. In the  $N=1$ case, only $m_1 \leq 1$ is allowed due to the square root in the second entry. In $N=2$ case the similar condition reads $m_1+m_2\leq1$, while in the $N=3$ case it is $m_1+m_2+m_3 \leq 1$. In other words, within the parameter $m$-space the regions corresponding to sensible solutions are $N$-simplexes.

For general case, the number of flavors is $N_F=\binom{N+2}{2}$. The mass matrix and the real part of the moduli matrix can be schematically written as (restoring generic $\tilde g \not= 1$)
\begin{align}
\label{eq:massmatrix} M & = -\mbox{diag}\Bigl\{\sum\limits_{j=0}^{k_1}m_j+\sum\limits_{j=0}^{k_2}m_j\Bigr\}/2\,, \\
\label{eq:modulimatrix} H_0 & = \Biggl\{\exp\left(\sum\limits_{j=0}^{k_1}R_j+\sum\limits_{j=k_1+1}^{k_2}R_j/2\right)\sqrt{2-\Bigl(\sum\limits_{j=k_1+1}^{k_2}\frac{m_j}{g v}\Bigr)^2}\Biggr\}\,,
\end{align}
where $k_1, k_2$ are whole numbers in the range $k_1\leq k_2 \in [0,N]$. The condition on $m_i$'s that ensures validity of the solution is
\begin{equation}\label{eq:maxcond3}
\sum\limits_{i=1}^{N}m_i \leq \sqrt{2}g v\,.
\end{equation}
Notice that this condition restrict the total tension $T = v^2\sum_i m_i$ to be  at most $T \leq\sqrt{2}g v^3$. This confirms, at lest for generic parameters, that no domain wall in \refer{eq:exsol} has a core (see discussion in Sec.~\ref{sec:crude}). 

In summary, the solution \refer{eq:exsol} describe a configuration of $N$  domain walls with tensions $T_i = v^2 m_i$ at (naive) positions $x_i$ given in Eq.~\refer{eq:exes2}, provided that \refer{eq:maxcond3} is satisfied. The minimal model, where the solution lives has $N_F= \binom{N+2}{2}$ flavors and mass matrix $M$ and moduli matrix $H_0$ given in \refer{eq:massmatrix}-\refer{eq:modulimatrix}, respectively.

Interestingly, we could view any $N>1$ solution in Eq.~\refer{eq:exsol} as a combination of $N-1$ building blocks $u_1^{(1)}(x)\equiv u_{II}^{3F}(x)$.
Indeed, we can rewrite Eqs.~\refer{eq:exsol2}-\refer{eq:exsol3} as
\begin{align}
u_1^{(2)}(x) & = 2\log\bigl(e^{u_{II}^{3F}(x; m_1)/2}+e^{u_{II}^{3F}(x; m_1+m_2)/2}-1\bigr)\,, \\
u_1^{(3)}(x) & = 2\log\bigl(e^{u^{N=2}(x)/2}+e^{u_{II}^{3F}(x; m_1+m_2+m_3)/2}-1\bigr) = 2\log\bigl(e^{u_{II}^{3F}(x; m_1)/2}\nonumber \\ & +e^{u_{II}^{3F}(x; m_1+m_2)/2}+e^{u_{II}^{3F}(x; m_1+m_2+m_3)/2}-2\bigr)\,.
\end{align}
And similar relations apply for higher $N$ cases. We can understand this as a kind of `non-linear' chains. However, if we rephrase everything in terms of $v(x) \equiv e^{u(x)/2}-1$ everything becomes linear too. In particular, the master equation turns into
\begin{equation}\label{eq:masterdual}
\frac{1}{g^2v^2}\bigl(\partial_x^2v+v\partial_x^2v-(\partial_x v)^2\bigr)= (v+1)^2 -\Omega_0(x)
\end{equation} 
and denoting $v_1^{(1)}(x) \equiv v_1(x) \equiv e^{u_{II}^{3F}(x)/2}-1$ we can see that solutions corresponding to $N=2$ and $N=3$ case are simply given as
\begin{align}
v_1^{(2)}(x) &= v_1(x; m_1)+v_1(x;m_1+m_2) \,, \\
v_1^{(3)}(x) &= v_1(x; m_1)+v_1(x;m_1+m_2) +v_1(x;m_1+m_2+m_3)\,.
\end{align} 
And analogously for higher $N$. 

The ansatz \refer{eq:exsol} is not the only one we can consider. We may also take
\begin{equation}
u_2^{(N)}(x) = 3\log F_N(x)\,,
\end{equation}
where $F_N$ is the same function as in \refer{eq:exsol}.
The corresponding source term is given as
\begin{equation}
\Omega_0 = F_N^3 +3(F_N^{\prime})^2F_N-3F_N^{\prime\prime} F_N^{2}\,.
\end{equation} 
The minimal model has $N_F = \binom{N+3}{3}$ flavors. 
This solution also describes a configuration of $N$ domain walls at (naive) positions $x_i$ of Eq.~\refer{eq:exes2}. However, the tension of each wall is now $T_i = 3v^2 m_i/2$ and the condition of validity is again
\begin{equation}
\sum\limits_{i=1}^{N}m_i \leq \sqrt{2}gv\,,
\end{equation}
which places an upper bound on the total tension $T\leq 3g v^3/\sqrt{2}$. 
The solutions $v_2^{(N)}(x)\equiv e^{u_2^{(N)}(x)/3}-1$ can also be written as chains of single-wall solution $v_2^{(1)}(x)\equiv v_2(x) = e^{u_I^{4F}(x)/3}-1$ in complete analogy to the previous results:
\begin{align}
v_2^{(2)}(x) &= v_2(x; m_1)+v_2(x;m_1+m_2) \,, \\
v_2^{(3)}(x) &= v_2(x; m_1)+v_2(x;m_1+m_2) +v_2(x;m_1+m_2+m_3)\,.
\end{align} 

The most generic configuration, however, can be achieved by combination of both:
\begin{equation}\label{eq:allsol}
u_3^{(N,\tilde N)}(x) = u_1^{(N)}(x)+u_2^{(\tilde N)}(x) = 2\log F_N(x)+ 3\log \tilde{F}_{\tilde{N}}(x)\,.
\end{equation}
This solution lives in a model with $N_F = \binom{N+2}{2}\binom{\tilde{N}+3}{3}$ flavors. It depicts a configuration of $N$ domain wall with tensions $T_i = v^2 m_i$ and $\tilde N$ domain walls with tensions $T_i = 3v^2 \tilde m_i/2$. The condition now reads
\begin{equation}\label{eq:gencond}
\Bigl(\sum\limits_{i=1}^{N}m_i\Bigr)^2+\Bigl(\sum\limits_{j=1}^{\tilde N}\tilde m_i\Bigr)^2 \leq 2g^2v^2\,.
\end{equation}
Interestingly, the solutions $v_3^{(N,\tilde N)} \equiv e^{u_3^{(N,\tilde N)}(x)}-1$ cannot be written as linear chains.

We believe that the class of functions \refer{eq:allsol} exhaust all exact solutions given as a logarithm of sum of exponentials.
This is the main result of this paper. 

\subsection{Exact multi-wall solutions up to $N_F=10$ flavors}

In this subsection we will investigate solutions $u_{1}^{(N)}$ for $N=1$, $N=2$ and $N=3$ in detail. We are not going to study any solutions in the families $u_2^{(N)}$ or $u_{3}^{(N)}$ for the sake of simplicity.

The main focus of our analysis is to find special relations between parameters $m_i$ and $R_i$, which reduce the given solution to another solution in the model with a lesser number of flavors. These reductions can be achieved in two complementary ways. First, we can make some entries of the moduli matrix \refer{eq:modulimatrix} to vanish. In $m$-space such instances corresponds to the (part of) the boundary of the $N$-simplex, within which the solutions are valid. The second way is to make some entries of the mass matrix \refer{eq:massmatrix} equal. As we will see, these relations represents special `cuts' of the $N$-simplex. If two such cuts intersect, further reduction occurs along this intersection, etc.
The biggest reduction within this approach occurs when all $m_i$'s are equal. In such case, the ansatz \refer{eq:exsol} is reduced to a degenerate $\mathbf{1}$ $N$-chain, described in the previous section. 

The combination of both ways leads to an intricate web of exact solutions, which we call the \emph{hierarchy}. We can view each solution in the family \refer{eq:exsol} as a special \emph{parent} solution, which gives many \emph{child} solutions via applying these reductions. Furthermore, each child can have their own children. Each reduction effectively represents a loss of one freedom, either the freedom to control tensions of individual walls (given by parameters $m_i$) or their mutual separations (parameters $R_i$). Solutions, which does not have any free parameters are \emph{irreducible} and they are terminating points of the hierarchy. 

Let us illustrate these general remarks on the concrete examples. Throughout this subsection, we will revive the practice, which we introduced below Eq.~\refer{eq:ex3}, that we decorate each new solution by the number of flavors in the superscript and by a roman numeral in the subscript (distinguishing solutions of the same flavor number) for ease of reference.

\subsubsection{$N=1$}

\begin{figure}[!h]
\includegraphics[width = 0.6\textwidth]{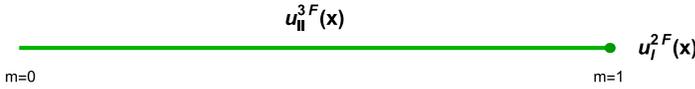}
\caption{\small Allowed region of mass parameter $m$ for $N=1$ case. The corresponding solution for each region is indicated.}
\label{fig:1w}
\end{figure}

Let us repeat the solution and $\Omega_0$ for reference:
\begin{equation}\label{eq:n2sol}
u_1^{(1)} \equiv u_{II}^{3F}(x) = 2 \log\bigl(1+e^{-m x}\bigr)\,, \hspace{5mm}
\Omega_0 = 1+2\bigl(1-m^2\bigr)e^{-m x}+e^{-2mx}\,.
\end{equation}
For simplicity we have denoted $m_1 \equiv m$ and set the moduli $R_1 = 0$. 
The mass matrix and (real part of) the moduli matrix reads
\begin{align}
M &= -\mbox{diag}\bigl(0\,, m\,, 2m\bigr)/2\,, \\
H_0 &= (1,\sqrt{2-2m^2},1)\,.
\end{align}
The allowed range of the parameter $m$ is the 1-simplex, the interval $[0,1]$. The (naive) positions of elementary walls are 
\begin{align}
x_1 & = \frac{1}{m}\log\bigl(2-2m^2\bigr)\,, \\
x_2 & = -\frac{1}{m}\log\bigl(2-2m^2\bigr)\,.
\end{align}
Within the 1-simplex $x_1\leq x_2$, so elementary walls are compressed into a single wall located at the origin (their center of mass).
The only reduction of a number of flavors  can be achieved by setting $m=1$, which makes the second entry in the moduli matrix zero. From the above formulas we can see that in this limit the compression becomes infinite $x_1\to-\infty$, $x_2\to\infty$ and the solution $u_{II}^{3F}(m=1) = u_{I}^{2F}$ can be identified with previously known 2F solution. We visualize these findings diagrammatically in Fig.~\ref{fig:1w}.

\subsubsection{$N=2$}

The solution and $\Omega_0$ reads: 
\begin{gather}\label{eq:n3sol}
u_1^{(2)} \equiv u_I^{6F}(x) = 2 \log\bigl(1+e^{R}e^{-m_1 x}+e^{-(m_1+m_2)x}\bigr)\,, \\
\Omega_0 =1+2e^{R}\bigl(1-m_1^2\bigr)e^{-m_1 x}
+e^{2R}e^{-2m_1 x}+2e^{R}e^{-(2m_1+m_2)x}\bigl(1-m_2^2\bigr)\nonumber \\
+2e^{-(m_1+m_2)x}\bigl(1-(m_1+m_2)^2\bigr)+e^{-2(m_1+m_2)x}\,. 
\end{gather}
We have set $R_2 = -R_1$ to put the center of mass at the origin and relabel $R_1 \equiv R$ for simplicity.

First, let us verify that the solution $u_I^{6F}(x)$ describes a configuration of two non-elementary wall located at 
\begin{equation}\label{eq:centers}
\tilde x_1 = \frac{R}{m_1}\,, \hspace{5mm} \tilde x_2 = -\frac{R}{m_2}\,.
\end{equation}
Contrary to the previous case the mass matrix cannot be ordered in the same way for all values of the parameters $m_1$ and $m_2$. In fact, there are two possible orderings:
{\small 
\begin{align}
M = & \, -\mbox{diag}\bigl(0\,,m_1\,, 2m_1\,, m_1+m_2\,, 2m_1+m_2\,, 2m_1+2m_2\bigr)/2\,, & m_2 &> m_1\,,\\
M = & \, -\mbox{diag}\bigl(0\,,m_1\,, m_1+m_2\,, 2m_1\,, 2m_1+m_2\,, 2m_1+2m_2\bigr)/2\,, & m_2 &\leq m_1\,.
\end{align}} 
The positions of elementary walls in the first case reads
\begin{align}
\label{eq:x1} x_1 &= \frac{1}{m_1}\Bigl[R+\log\bigl(2-2m_1^2\bigr)/2\Bigr]\,, \\
\label{eq:x2} x_2 &= \frac{1}{m_1}\Bigl[R-\log\bigl(2-2m_1^2\bigr)/2\Bigr]\,, \\
\label{eq:x3} x_3 & = \frac{1}{m_2-m_1}\Bigl[-2R+\log\bigl(2-2(m_1+m_2)^2\bigr)/2\Bigr]\,,  \\
\label{eq:x4} x_4 & = \frac{1}{m_1}\Bigl[R+\log\Bigl(\frac{1-m_2^2}{1-(m_1+m_2)^2}\Bigr)/2\Bigr]\,,  \\
\label{eq:x5} x_5 & = \frac{1}{m_2}\Big[-R-\log\bigl(2-2m_2^2\bigr)/2\Bigr]\,.
\end{align}
The allowed range of the parameters $m_1$ and $m_2$ is a 2-simplex (a rectangular triangle) defined by the relation $m_1+m_2 \leq 1$. This means that $m_1\leq1$ and hence $x_1\leq x_2$. Thus, first two elementary walls form a compressed wall located at their center of mass $\tilde x_1 = R/m_1$ as we claimed. Assuming that $R>0$ we see that $x_4 >0$ and $x_3<0$, which means that third and fourth wall are compressed with each other. Their center of mass is located at $-R/m_2+\log\bigl(2-2m_2^2\bigr)/2$ which is always smaller than $x_5$, indicating that fifth wall is compressed as well. Their total center of mass lies at $\tilde x_2 = -R/m_2$, in confirmation with Eq.~\refer{eq:centers}. 

On the other hand, in the second case $m_2 \leq m_1$, the positions of elementary walls are given as follows
 \begin{align}
\label{eq:xx1} x_1 &= \frac{1}{m_1}\Bigl[R+\log\bigl(2-2m_1^2\bigr)/2\Bigr]\,, \\
\label{eq:xx2} x_2 &= \frac{1}{m_2}\Bigl[-R+\log\Bigl(\frac{1-(m_1+m_2)^2}{1-m_1^2}\Bigr)/2\Bigr]\,, \\
\label{eq:xx3} x_3 & = \frac{1}{m_1-m_2}\Bigl[2R-\log\bigl(2-2(m_1+m_2)^2\bigr)/2\Bigr]\,,  \\
\label{eq:xx4} x_4 & = \frac{1}{m_2}\Bigl[-R+\log\bigl(2-2m_2^2\bigr)/2\Bigr]\,,  \\
\label{eq:xx5} x_5 & = \frac{1}{m_2}\Big[-R-\log\bigl(2-2m_2^2\bigr)/2\Bigr]\,.
\end{align}
The analysis is completely analogous but now the roles are reversed. The last two walls are compressed  with each other (since $x_4<x_5$) and the first three as well (since for $R>0$ we have $x_3>0$, $x_2<0$ and $(m_1+m_2)x_3+m_2 x_2>m_1 x_1$). The center of mass of the triplet is located at $\tilde x_1 = R/m_1$, while that of the doublet is at $\tilde x_2 = -R/m_2$, again with full accordance to Eq.~\refer{eq:centers}.

It is clear that for higher $N$ the positions of the elementary wall will be increasingly laborious to ascertain due to the ever higher number of possible orderings of the elements in the mass matrix $M$. The previous analysis, however, confirmed that we can be confident of locations of compressed walls given in Eq.~\refer{eq:exes2}, which we extracted directly from the solution \refer{eq:exsol}. 

\begin{figure}[!h]
\includegraphics[width = 0.9\textwidth]{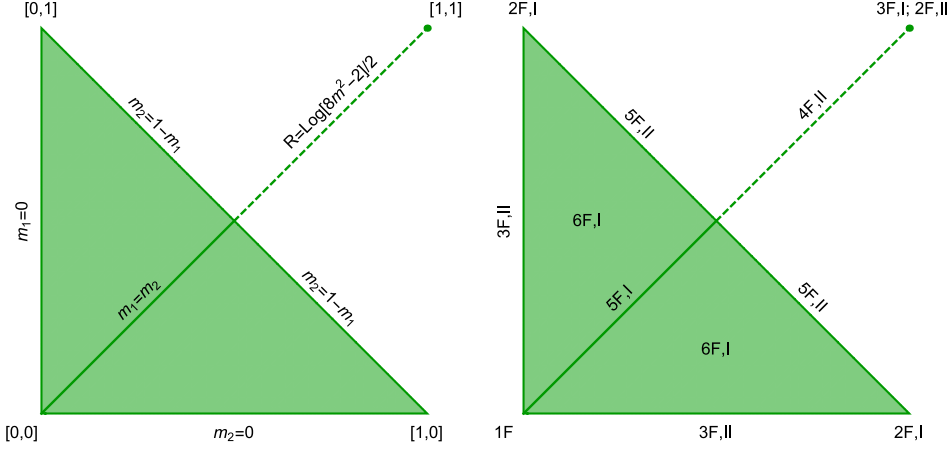}
\caption{\small Allowed region of mass parameters $m_1$ and $m_2$ for $N=2$ case. In the left part the edges and vertices are specified, while in the right we show which solution each segment represents. The points are formatted as $[m_1,m_2]$.}
\label{fig:2w}
\end{figure}

On Fig.~\ref{fig:2w} we show the 2-simplex in two versions. In first, we  show relations among $m_i$'s that define its boundary and in the second we put a label on each face, edge and vertex specifying which solution a given region represents.
This figure summarize where reductions of the number of flavors occur. As discussed in the previous subsection, sources of reduction are twofold. The first is where some entries of the moduli matrix $H_0$ vanishes. There are only three possibilities, which defines the boundary of the triangle.
The edges $m_1 = 0$ and $m_2 = 0$ returns us to the $N=1$ case, therefore we already know the answer there. The last edge of the triangle $m_2 = 1-m_1$ represents a new 5F solution:
\begin{align}
u_{II}^{5F}(x) & = 2\log\bigl(1+e^{R}e^{-m x}+e^{-x}\bigr)\,, \\
\Omega_0 & = 1+2e^{R}e^{-m x}(1-m^2)+e^{2R}e^{-2m x}+2e^{R}e^{-(1+m)x}m(2-m)+e^{-2x}\,,
\end{align}
where we have relabelled the parameters $m_1=m$ and $m_2 = 1-m$ for simplicity. This solution can be reduced further by setting $m=1$ or $m=0$, which leads to the $u_{I}^{2F}(x)$ solution.
 
The second way how to reduce the number of flavors is to make some elements of the mass matrix $M$ identical. In the present case, there is only one interesting possibility: $m_1 = m_2 \equiv m$. This is the 2-chain, which we encountered in Eq.~\refer{eq:5Fsol}:
\begin{align}\label{eq:5Fsol2}
u_{I}^{5F}(x) & = 2\log\bigl(1+e^{R}e^{-m x}+e^{-2mx}\bigr) = u_{II}^{3F}(x+S)+u_{II}^{3F}(x-S)\,, \\
\Omega_0 & = 1+2e^{R}e^{-m x}(1-m^2)+e^{-2m x}\bigl(2-8m^2+e^{2R}\bigr)+2e^{R}e^{-3mx}\bigl(1-m^2\bigr)\nonumber \\&+e^{-4mx}\,, \label{eq:omega02} 
\end{align}
where $S = \mbox{arccosh}(R/2)/m$.
The curious property of this solution is that the parameter $m$ can escape the 2-simplex and still give a well-defined solution for arbitrary separation of walls. But when $m\geq 1/2$ it is preferable to redefine parameter $R$ as
\begin{equation}
e^{R}\to \sqrt{e^{2R}+8m^2-2}\,,
\end{equation}
so that every coefficient in $\Omega_0$ is manifestly positive.
We can reduce the number of flavors further by setting $m=1$, which produce a known $3F$ solution $u_{I}^{5F}(m=1) = u_{I}^{3F}$.

\begin{figure}[!h]
\includegraphics[width = 0.5\textwidth]{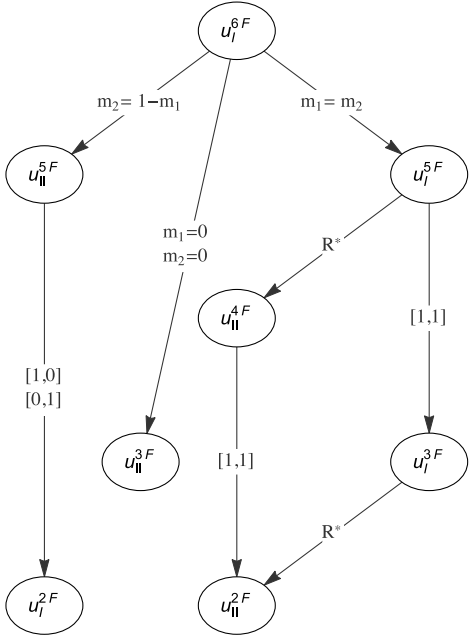}
\caption{\small A schematic representation of reductions of $u_I^{6F}(x)$ solution. The labels follow the structure of Fig.~\ref{fig:2w}. $R^{*}$ indicates that special choice of positions of domain walls is taken.}
\label{fig:n2tree}
\end{figure}

There is, however, still one option left. We can also reduce the number of flavors of $u_{I}^{5F}(x)$ by fixing the positions of the wall to specific values, which will make the third factor in Eq.~\refer{eq:omega02} disappear. In other words, we set $R=\log(8m^2-2)/2$. This can be done only for $m\in [1/2,1]$ and the resulting solution is
 \begin{align}
u_{II}^{4F}(x) & = 2\log\bigl(1+\sqrt{8m^2-2}e^{-m x}+e^{-2mx}\bigr)\,, \\
 \Omega_0 & = 1+2e^{-m x}\sqrt{8m^2-2}(1-m^2)+2e^{-3mx}\sqrt{8m^2-2}\bigl(1-m^2\bigr)+e^{-4mx}\,.
\end{align}
In Fig.~\ref{fig:2w} we have indicated this special solution by a dashed line.
Lastly, by setting $m=1$ in $u_{II}^{4F}(x)$ we obtain $u_{II}^{4F}(m=1) = u_{II}^{2F}$.
With this observation, we have exhausted all possible reductions in $N=2$ case.
We summarize the structure of reductions on Fig.~\ref{fig:n2tree}. 

\subsubsection{$N=3$}

The solution and $\Omega_0$ reads: 
\begin{align}\label{eq:n4sol}
u_1^{(3)} & \equiv u_I^{10F}(x) = 2 \log\bigl(1+e^{R_1}e^{-m_1 x}+e^{R_1+R_2}e^{-(m_1+m_2)x}+e^{-(m_1+m_2+m_3) x}\bigr)\,, \\
\label{eq:omega03n} \Omega_0 &= 1+2e^{R_1}e^{-m_1x}\bigl(1-m_1^2\bigr)+2e^{R_1+R_2}e^{-(m_1+m_2)x}\bigl(1-(m_1+m_2)^2\bigr)\nonumber \\
& +2e^{-(m_1+m_2+m_3)x}\bigl(1-(m_1+m_2+m_3)^2\bigr)+2e^{2R_1+R_2}e^{-(2m_1+m_2)x}\bigl(1-m_2^2\bigr)
\nonumber \\
& +2e^{R_1}e^{-(2m_1+m_2+m_3)x}\bigl(1-(m_2+m_3)^2\bigr)+2e^{R_1+R_2}e^{-(2m_1+2m_2+m_3)x}\bigl(1-m_3^2\bigr) 
\nonumber \\
&+e^{2R_1+2R_2}e^{-(2m_1+2m_2)x}+e^{-2(m_1+m_2+m_3)x}+e^{2R_1}e^{-2m_1 x}\,. 
\end{align}
We have set $R_3 = -R_1-R_2$ to put the center of mass at the origin. The solution $u_{I}^{10F}(x)$ describes three domain walls located at
\begin{equation}\label{eq:exes3}
\tilde x_1 = \frac{R_1}{m_1}\,, \hspace{5mm}
\tilde x_2 = \frac{R_2}{m_2}\,, \hspace{5mm}
\tilde x_3 = -\frac{R_1+R_2}{m_3}\,.
\end{equation}

\begin{figure}[!h]
\includegraphics[width = 0.9\textwidth]{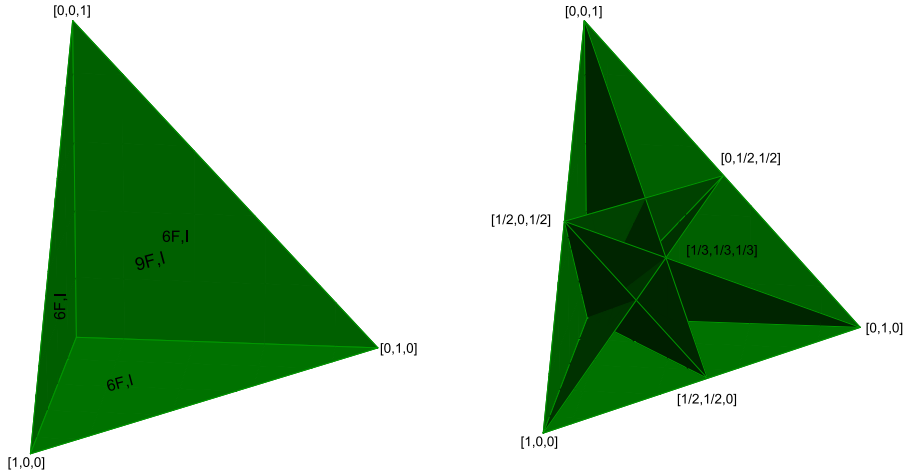}
\caption{\small The allowed parameter region for $N=3$ case is a rectangular tetrahedron. In the left panel, we see the generic solutions attached to each face. Faces where one of the parameters vanishes corresponds to $N=2$ case, while the top face is a new 9F solution. In the right panel, we `unlit' the top face to reveal the inner structure of the tetrahedron, which is cut by five planes. These cuts represent mass matrix degeneracies of the general solution. The points shown on the pictures are formatted as $[m_1,m_2,m_3]$.}
\label{fig:3w}
\end{figure}

The allowed region in $m$-space is a rectangular tetrahedron $m_1+m_2+m_3 \leq 1$ (see Fig.~\ref{fig:3w}). The boundary of this 3-simplex is where some entries of the moduli matrix vanishes.  
The faces that are attached to the origin -- rectangular triangles $m_1 =0$, $m_2=0$ and $m_3=0$ -- are where we return to the $N=2$ case. 

\begin{figure}[!h]
\includegraphics[width = 0.9\textwidth]{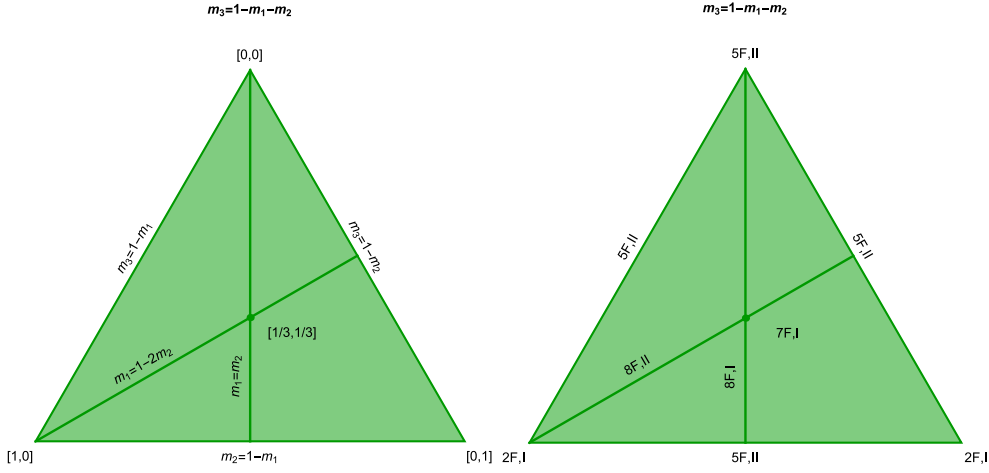}
\caption{\small Allowed region of mass parameters $m_1$ and $m_2$ for $N=3$ case for the face of the tetrahedron defined by the relation $m_3 = 1-m_1-m_2$. In the left part, the edges and vertices are specified, while in the right we show which solution each segment represents. The points are formatted as $[m_1,m_2]$.}
\label{fig:3wface}
\end{figure}

The remaining face $m_3 = 1-m_1-m_2$ corresponds  to the first 9F solution:
\begin{align}
u_I^{9F}(x)  &= 2 \log\bigl(1+e^{R_1}e^{-m_1 x}+e^{R_1+R_2}e^{-(m_1+m_2)x}+e^{- x}\bigr)\,, \\
\Omega_0 &= 1+2e^{R_1}e^{-m_1x}\bigl(1-m_1^2\bigr)+2e^{R_1+R_2}e^{-(m_1+m_2)x}\bigl(1-(m_1+m_2)^2\bigr)\nonumber \\
& +2e^{2R_1+R_2}e^{-(2m_1+m_2)x}\bigl(1-m_2^2\bigr)+2e^{R_1}e^{-(1+m_1)x}m_1(2-m_1)+e^{-2x}
\nonumber \\
& +2e^{R_1+R_2}e^{-(1+m_1+m_2)x}(m_1+m_2)(2-m_1-m_2) +e^{2R_1+2R_2}e^{-(2m_1+2m_2)x}
\nonumber \\
& +e^{2R_1}e^{-2m_1 x}\,.
\end{align}
We can further lower the number of flavors of this solution by investigation linear relations among parameters $m_1$ and $m_2$. The full structure of reductions is displayed in Fig.~\ref{fig:3wface}. 

The relations 
\begin{equation}
m_1 = m_2\,, \hspace{5mm} m_1 = 1-2m_2\,,
\end{equation}
give us new 8F solutions:
\begin{align}
u_{I}^{8F}(x) & = 2\log\bigl(1+e^{R_1}e^{-m x}+e^{R_1+R_2}e^{-2m x}+e^{-x}\bigr)\,, \\
\Omega_0 & = 1+2e^{-m x}e^{R_1}\bigl(1-m^2\bigr)+e^{-2m x}e^{R_1}\bigl(e^{R_1}+2e^{R_2}(1-4m^2)\bigr)
\nonumber \\
& +2e^{-3m x}e^{2R_1+R_2}\bigl(1-m^2\bigr)+2e^{-(1+m)x}e^{R_1}m(2-m) 
\nonumber \\
& +e^{-4m x}e^{2R_1+2R_2}+8e^{-(1+2m)x}e^{R_1+R_2}m(1-m)+e^{-2x}\,. \\
& \nonumber \\
u_{II}^{8F}(x) & = 2\log\bigl(1+e^{R_1}e^{-(1-2m) x}+e^{R_1+R_2}e^{-(1-m) x}+e^{-x}\bigr)\,, \\
\Omega_0 & = 1+8e^{-(1-2m) x}e^{R_1}m(1-m)+2e^{-(1-m) x}e^{R_1+R_2}m(2-m)
\nonumber \\
& +2e^{-(2-3m)x}e^{2R_1+R_2}\bigl(1-m^2\bigr)+2e^{-(2-m)x}e^{R_1+R_2}\bigl(1-m^2\bigr) 
\nonumber \\
& +e^{-2(1-m)x}e^{R_1}\bigl(e^{R_1+2R_2}+2-8m^2\bigr)+e^{-2(1-2m)x}e^{2R_1}+e^{-2x}\,,
\end{align}
where for $u_I^{8F}(x)$ solution we relabelled $m_1 = m_2 \equiv m$ and for $u_{II}^{8F}(x)$ solution we set $m_2 \equiv m$, $m_1 = 1-2m $ for simplicity. Although different solutions, these two functions are related via a `reflection' identity
\begin{equation}
u_I^{8F}(x;R_1,R_2) = u_{II}^{8F}(-x;R_1+R_2,-R_2)-2x\,.
\end{equation}
Also notice that the range of the parameter $m$ for both $u_{I}^{8F}(x)$, $u_{II}^{8F}(x)$ is $m\in [0,1/2]$. The reason is that inside the interval $m\in [1/2,1]$ both solutions are unbalanced (in the sense as described below Eq.~\refer{eq:has}). If we rebalance them we obtain:
\begin{align}
u_{III}^{8F}(x) & = 2\log\bigl(1+e^{R_1}e^{-m x}+e^{R_1+R_2}e^{-x}+e^{-2m x}\bigr)\,, \\
u_{IV}^{8F}(x) & = 2\log\bigl(1+e^{R_1}e^{-(2m-1) x}+e^{R_1+R_2}e^{-m x}+e^{-2 m x}\bigr)\,.
\end{align}
These new solutions are connected with $u_{I}^{8F}(x)$, $u_{II}^{8F}(x)$ via the transformations
{\small \begin{align}
u_{III}^{8F}(x; R_1, R_2) &= u_I^{8F}\bigl(x-R_1-R_2; (1+m)R_1-mR_2,-(1+m)R_1-mR_2\bigr)\,, \\
u_{IV}^{8F}(x; R_1, R_2) &= u_{II}^{8F}\bigl(x+R_1; -2mR_1, R_2+(3m-1)R_1\bigr)-2(2m-1) x+2R_1\,.
\end{align}}
However, solutions $u_{III}^{8F}(x)$ and $u_{IV}^{8F}(x)$ are obviously not a part of the top face, since for both $m_3 \not = 1-m_1-m_2$. In fact, they are situated on planes $m_1 = m_2+m_3$ and $m_3=m_1+m_2$ respectively, which are inside the tetrahedron and corresponds to reductions of mass matrix elements. We will discuss such planes shortly.
Both solutions $u_{III}^{8F}(x)$ and $u_{IV}^{8F}(x)$  can be reduced by one flavor by taking $R_1 = \log\bigl(8m^2-2\bigr)/2$ and $R_2 = -R_1 +\bigl(8m^2-2\bigr)/2$, respectively. These  new 7F solutions
\begin{align}
u_{II}^{7F}(x) & = 2\log\bigl(1+\sqrt{8m^2-2}e^{-m x}+e^{R_2}\sqrt{8m^2-2}e^{-x}+e^{-2m x}\bigr)\,, \\
u_{III}^{7F}(x) & = 2\log\bigl(1+e^{R_1}e^{-(2m-1) x}+\sqrt{8m^2-2}e^{-m x}+e^{-2 m x}\bigr)\,,
\end{align} 
cannot be reduced further.

At  the point on the top face $m_1 =m_2= 1/3$, where the two 8F solutions $u_I^{8F}$ and $u_{II}^{8F}$ meet, the number of flavors is reduced by one. This is not, however, a new solution, rather it is a special point of $u_{I}^{7F}(x)$ solution given in Eq.~\refer{eq:7Fsol}.

With this observation, we have finished the analysis of the upper face of the tetrahedron and we can now move to explore relations between mass matrix elements. These are
\begin{equation}
m_1 = m_2\,, \hspace{5mm} m_1 = m_2+m_3 \,, \hspace{5mm}
m_1 = m_3 \,, \hspace{5mm} m_3 = m_1+m_2\,, \hspace{5mm}
m_2 = m_3\,.
\end{equation}
Above relations defines planes, which cuts the rectangular tetrahedron. Along these cuts, the solution is reduced by one flavor. We display these cuts on Fig.~\ref{fig:3w}. The corresponding solutions are (in order)
{\small \begin{align}
u_{II}^{9F}(x; m_1,m_3) & = 2\log\bigl(1+e^{R_1}e^{-m_1 x}+e^{R_1+R_2}e^{-2m_1 x}+e^{-(2m_1+m_3) x}\bigr)\,, \\ 
u_{III}^{9F}(x; m_2,m_3) & = 2\log\bigl(1+e^{R_1}e^{-(m_2+m_3) x}+e^{R_1+R_2}e^{-(2m_2+m_3) x}+e^{-2(m_2+m_3) x}\bigr)\,, \\ 
u_{IV}^{9F}(x; m_1,m_2) & = 2\log\bigl(1+e^{R_1}e^{-m_1 x}+e^{R_1+R_2}e^{-(m_1+m_2) x}+e^{-(2m_1+m_2) x}\bigr)\,, \\ 
u_{V}^{9F}(x; m_1,m_2) & = 2\log\bigl(1+e^{R_1}e^{-m_1 x}+e^{R_1+R_2}e^{-(m_1+m_2) x}+e^{-2(m_1+m_2) x}\bigr)\,, \\ 
u_{VI}^{9F}(x;m_1,m_2) & = 2\log\bigl(1+e^{R_1}e^{-m_1 x}+e^{R_1+R_2}e^{-(m_1+m_2) x}+e^{-(m_1+2m_2) x}\bigr)\,.
\end{align}}
Not all of these solutions are independent. Solutions  and $u_{II}^{9F}(x)$, $u_{III}^{9F}(x)$ and $u_{V}^{9F}(x)$ are related by redefinition of parameters
\begin{equation}
u_{III}^{9F}(x;m_1+m_3,-m_3) \sim u_{V}^{9F}(x;2m_1+m_3,-m_1-m_3) \sim u_{II}^{9F}(x;m_1,m_3)\,,  
\end{equation}
where $\sim$ means up to inconsequential rebalancing involving shifts and redefinitions of parameters $R_1$ and $R_2$. Furthermore, solution $u_{VI}^{9F}(x)$ is related to $u_{II}^{9F}(x)$ by another reflection identity
\begin{equation}\label{eq:refl}
u_{VI}^{9F}(-x;m_3,m_1) -2(m_3+2m_1)x \sim u_{II}^{9F}(x; m_1,m_2)\,.
\end{equation}
Notice that both $u_{I}^{9F}(x)$ and $u_{IV}^{9F}(x)$ are self-dual under reflection:
\begin{align}
u_{I}^{9F}(-x;1-m_1,-m_2) -2x  & = u_{I}^{9F}(x;m_1,m_2)\,, \\
u_{IV}^{9F}(-x;m_1,m_2) -2(2m_1+m_2)x & \sim u_{IV}^{9F}(x; m_1,m_2)\,,
\end{align}
Given these identities we need to closely examine only, say, $u_{II}^{9F}(x)$, $u_{VI}^{9F}(x)$ and $u_{IV}^{9F}(x)$.

\begin{figure}[!h]
\includegraphics[width = 1.0\textwidth]{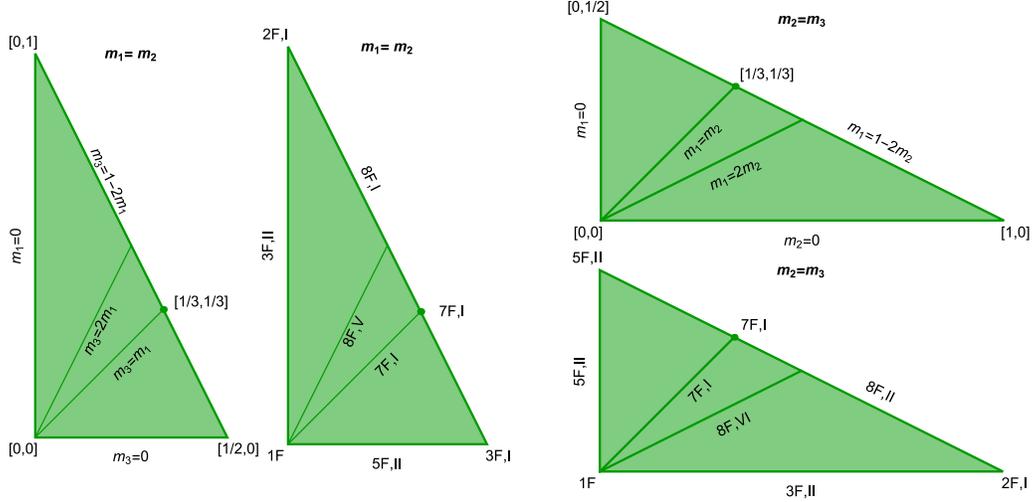}
\caption{\small Cuts of the tetrahedron defined by the relations $m_1 = m_2$ and $m_2 = m_3$.}
\label{fig:planes}
\end{figure}

In Fig.~\ref{fig:planes} we display the structure of solutions $u_{II}^{9F}(x)$ and $u_{VI}^{9F}(x)$. There are two new 8F solutions
\begin{align}
u_{V}^{8F}(x) & = 2\log\bigl(1+e^{R_1}e^{-m x}+e^{R_1+R_2}e^{-2mx}+e^{-4mx}\bigr)\,, \\
u_{VI}^{8F}(x) & = 2\log\bigl(1+e^{R_1}e^{-2m x}+e^{R_1+R_2}e^{-3mx}+e^{-4mx}\bigr)\,,
\end{align}
where in the first one $m_1 = m_2 = m_3/2 \equiv m$ and in the second one $m_1/2 = m_2 = m_3 \equiv m$.
They are related to each other via reflection identity
\begin{equation}
u_{V}^{8F}(-x)-8mx \sim u_{VI}^{8F}(x)\,,
\end{equation}
which is inherited from the more general identity in Eq.~\refer{eq:refl}.
Both of these solutions can be reduced by one flavor, either by taking $m=1/3$ or assuming special positions of domain walls, namely $R_1 = -R_2 +\log\bigl(32m^2-2\bigr)/2$ for $u_{V}^{8F}(x)$ and $R_1 =\log\bigl(32m^2-2\bigr)/2 $ in the case of $u_{V}^{8F}(x)$.  
Both approaches gives us new 7F solutions:
\begin{align}
u_{IV}^{7F}(x) & = 2\log\bigl(1+e^{-R_2}\sqrt{32m^2-2}e^{-m x}+\sqrt{32m^2-2}e^{-2mx}+e^{-4mx}\bigr)\,, \\
u_{V}^{7F}(x) & = 2\log\bigl(1+e^{R_1}e^{-x/3}+e^{R_1+R_2}e^{-2x/3}+e^{-4x/3}\bigr)\,, \\
u_{VI}^{7F}(x) & = 2\log\bigl(1+e^{R_1}\sqrt{32m^2-2}e^{-2m x}+e^{-R_1}\sqrt{32m^2-2}e^{-3mx}+e^{-4mx}\bigr)\,, \\
u_{VII}^{7F}(x) & = 2\log\bigl(1+e^{R_1}e^{-2x/3}+e^{R_1+R_2}e^{-x}+e^{-4x/3}\bigr)\,.
\end{align} 
Furthermore, taking both $m=1/3$ and special positions of domain walls yields new 6F solutions:
\begin{align}
u_{II}^{6F}(x) & = 2\log\bigl(1+e^{-R_2}\sqrt{14/9}e^{- x/3}+\sqrt{14/9}e^{-2x/3}+e^{-4x/3}\bigr)\,, \\
u_{III}^{6F}(x) & = 2\log\bigl(1+e^{R_1}\sqrt{14/9}e^{-2x/3}+e^{-R_1}\sqrt{14/9}e^{-x}+e^{-x/3}\bigr)\,,
\end{align}
which cannot be reduced further.

As we can see in Fig.~\ref{fig:3w} the planes $m_1=m_2$ and $m_2=m_3$ intersect at a line $m_1=m_2=m_3$, which gives 7F solution
\begin{align}\label{eq:7Fsol}
u_{I}^{7F}(x) &= 2\log\bigl(1+e^{R_1}e^{-mx}+e^{R_1+R_2}e^{-2mx}+e^{-3mx}\bigr)\,, \\
\Omega_0 & = 1+2e^{R_1}e^{-m x}\bigl(1-m^2\bigr)+e^{-2mx}\bigl(e^{2R_1}+2e^{R_1+R_2}(1-4m^2)\bigr)
\nonumber \\ & +2e^{-3mx}\bigl(e^{2R_1+R_2}(1-m^2)+1-9m^2\bigr)+e^{-4mx}\bigl(e^{2R_1+2R_2}+2e^{R_1}(1-4m^2)\bigr)
\nonumber \\ & +2e^{R_1+R_2}e^{-5mx}\bigl(1-m^2\bigr)+e^{-6mx}\,.
\end{align}
This solution is a degenerate 3-chain
\begin{equation}
u_{I}^{7F}(x) = u_{II}^{3F}(x-S_1)+u_{II}^{3F}(x-S_2)+u_{II}^{3F}(x+S_1+S_2)\,,
\end{equation} 
where
\begin{equation}
e^{R_1} = e^{m S_1}+e^{m S_2}+e^{-m (S_1+S_2)}\,, \hspace{5mm}
e^{R_1+R_2} = e^{-m S_1}+e^{-m S_2}+e^{m (S_1+S_2)}\,.
\end{equation}
We cannot reduce this solution further unless we fix the position parameters $R_1, R_2$. This can be done in several ways. If  $m>1/3$ then we can fix
\begin{equation}
R_2 = -2R_1 +\log\frac{9m^2-1}{1-m^2}\,,
\end{equation} 
to get a 6F solution
\begin{equation}
u_{IV}^{6F}(x) = 2\log\bigl(1+e^{R_1}e^{-mx}+\tfrac{9m^2-1}{1-m^2}e^{-R_1}e^{-2mx}+e^{-3mx}\bigr)\,.
\end{equation}
 Furthermore, if $m>1/2$ we can fix $R_1$ in two ways
\begin{align}
R_1 & = R_2 +\log\bigl(8m^2-2\bigr)\,, \\
R_1 & =  -2R_2+ \log\bigl(8m^2-2\bigr)\,.
\end{align}
Each choice reduce the solution by one flavor:
\begin{align}
u_{V}^{6F}(x) &= 2\log\bigl(1+e^{R_2}e^{-mx}\bigl(8m^2-2\bigr)+e^{2R_2}e^{-2mx}\bigl(8m^2-2\bigr)+e^{-3mx}\bigr)\,, \\
u_{VI}^{6F}(x) &= 2\log\bigl(1+e^{-2R_2}e^{-mx}\bigl(8m^2-2\bigr)+e^{-R_2}e^{-2mx}\bigl(8m^2-2\bigr)+e^{-3mx}\bigr)\,.
\end{align} 
We can combine the above relations and  fix both $R_1$ and $R_2$ in three independent ways  
\begin{align}
R_1 & = \frac{1}{3}\log \frac{(9m^2-1)(8m^2-2)}{1-m^2}\,, & R_2 &= \frac{1}{3}\log \frac{9m^2-1}{(1-m^2)(8m^2-2)^2}\,, \\
R_1 & = \frac{1}{3}\log \frac{(9m^2-1)^2}{(1-m^2)^2(8m^2-2)}\,, & R_2 &= \frac{1}{3}\log \frac{(1-m^2)(8m^2-2)^2}{9m^2-1}\,, \\
R_1 & = \log(8m^2-2)\,, & R_2 &= 0\,,
\end{align}
with corresponding 5F solutions {\small
\begin{align}
u_{III}^{5F}(x) & = 2\log\bigl(1+\sqrt[3]{\tfrac{(9m^2-1)(8m^2-2)}{1-m^2}}e^{-mx}+\sqrt[3]{\tfrac{(9m^2-1)2}{(1-m^2)^2(8m^2-2)}}e^{-2mx}+e^{-3mx}\bigr)\,, \\
u_{IV}^{5F}(x) &= 2\log\bigl(1+\sqrt[3]{\tfrac{(9m^2-1)^2}{(1-m^2)^2(8m^2-2)}}e^{-mx}+\sqrt[3]{\tfrac{(9m^2-1)(8m^2-2)}{1-m^2}}e^{-2mx}+e^{-3mx}\bigr)\,, \\
u_{V}^{5F}(x) &= 2\log\bigl(1+(8m^2-2)e^{-mx}+(8m^2-2)e^{-2mx}+e^{-3mx}\bigr)\,.
\end{align}}
And finally if we choose special values for $m$ for which all three relations are satisfied we obtain most flavor reduced solutions. These special values are the roots of the equation $(8m^2-1)^2(1-m^2)=9m^2-1$. It turns out that there are two for which $m\in [1/2,1]$. Let us denote these roots as $\tilde m$ and $\bar m$. Their approximate values are  $\tilde m \approx 0.883$ and $\bar m \approx 0.597$. If we choose the first root we obtain $R_1 =  \log(8\tilde m^2-2) \approx 1.444$, $R_2 = 0$ and the corresponding solution is given as 
\begin{align}
u_{III}^{4F}(x) & = 2\log\bigl(1+(8\tilde m^2-2)e^{-\tilde m x}+(8\tilde m^2-2)e^{-2\tilde mx}+e^{-3\tilde mx}\bigr)\,, \\
\Omega_0 & = 1+2(8\tilde m^2-2)(1-\tilde m^2)e^{-\tilde m x}\bigl(1+e^{-4\tilde m x}\bigr)+e^{-6\tilde m x}\,.
\end{align}
For the other root we have $R_1 = \log(8\bar m^2-2) \approx 0.616$ and  $R_2 = 0$. The corresponding solution $u_{IV}^{4F}(x)$ is functionally the same as $u_{III}^{4F}(x)$ with $\tilde m$ replaced by $\bar m$.

\begin{figure}[!h]
\includegraphics[width = 0.9\textwidth]{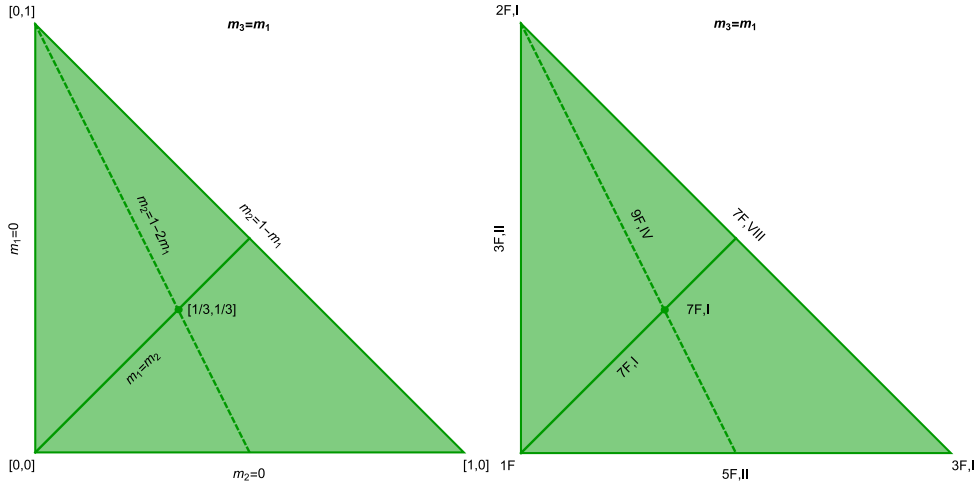}
\caption{\small Cut of the tetrahedron defined by the relation $m_3 = m_1$.}
\label{fig:plane3}
\end{figure}

Lastly, let us comment on the reduction structure of the solution $u_{IV}^{9F}(x)$, which is shown in Fig.~\ref{fig:plane3}.
The solution and corresponding $\Omega_0$ reads
\begin{align}
u_{IV}^{9F}(x) & = 2\log\bigl(1+e^{R_1}e^{-m_1 x}+e^{R_1+R_2}e^{-(m_1+m_2) x}+e^{-(2m_1+m_2) x}\bigr)\,, \\
\Omega_0 & = 1+2e^{R_1}e^{-m_1 x}\bigl(1-m_1^2\bigr)+2e^{R_1+R_2}e^{-(m_1+m_2)x}\bigl(1-(m_1+m_2)^2\bigr)
\nonumber \\ & +2e^{-(2m_1+m_2)x}\Bigl[e^{2R_1+R_2}\bigl(1-m_2^2\bigr)+1-\bigl(2m_1+m_2\bigr)^2\Bigr]+e^{2R_1+2R_2}e^{-2(m_1+m_2)x} \nonumber \\
& +2e^{R_1}e^{-(3m_1+m_2)x}\bigl(1-(m_1+m_2)^2\bigr)+2e^{R_1+R_2}e^{-(3m_1+2m_2)x}\bigl(1-m_1^2\bigr)\nonumber \\ &+e^{-2(2m_1+m_2)x}\,.
\end{align}
The values of parameters, which produce meaningful solutions, are constrained as $m_1+m_2 \leq 1$, but for $m_2\geq 1-2m_1$ we see that $R_2 \geq R^{*}\equiv -2R_1 +\log \bigl((2m_1+m)^2-1)/(1-m_2^2)\bigr)$ must hold so that the fourth term in $\Omega_0$ is non-negative. We indicate that by a dashed line in Fig.~\ref{fig:plane3}.
If $R_2 = R^{*}$ we obtain a new 8F solution
\begin{align}
u_{VII}^{8F}(x) & = 2\log\bigl(1+e^{R_1}e^{-m_1 x}+e^{-R_1}e^{-(m_1+m_2) x}\tfrac{(2m_1+m_2)^2-1}{1-m_2^2}+e^{-(2m_1+m_2) x}\bigr)\,, \\
\Omega_0 & = 1+2e^{R_1}e^{-m_1 x}\bigl(1-m_1^2\bigr)+2e^{-R_1}e^{-(m_1+m_2)x}\bigl(1-(m_1+m_2)^2\bigr)\tfrac{(2m_1+m_2)^2-1}{1-m_2^2}
\nonumber \\ &+e^{-2R_1}e^{-2(m_1+m_2)x}\tfrac{\bigl((2m_1+m_2)^2-1\bigr)^2}{(1-m_2^2)^2}+2e^{R_1}e^{-(3m_1+m_2)x}\bigl(1-(m_1+m_2)^2\bigr) \nonumber \\
& +2e^{-R_1}e^{-(3m_1+2m_2)x}\bigl(1-m_1^2\bigr)\tfrac{(2m_1+m_2)^2-1}{1-m_2^2}+e^{-2(2m_1+m_2)x}+e^{2R_1}e^{-2m_1 x}\,.
\end{align}

\begin{figure}[!h]
\includegraphics[width = 0.9\textwidth]{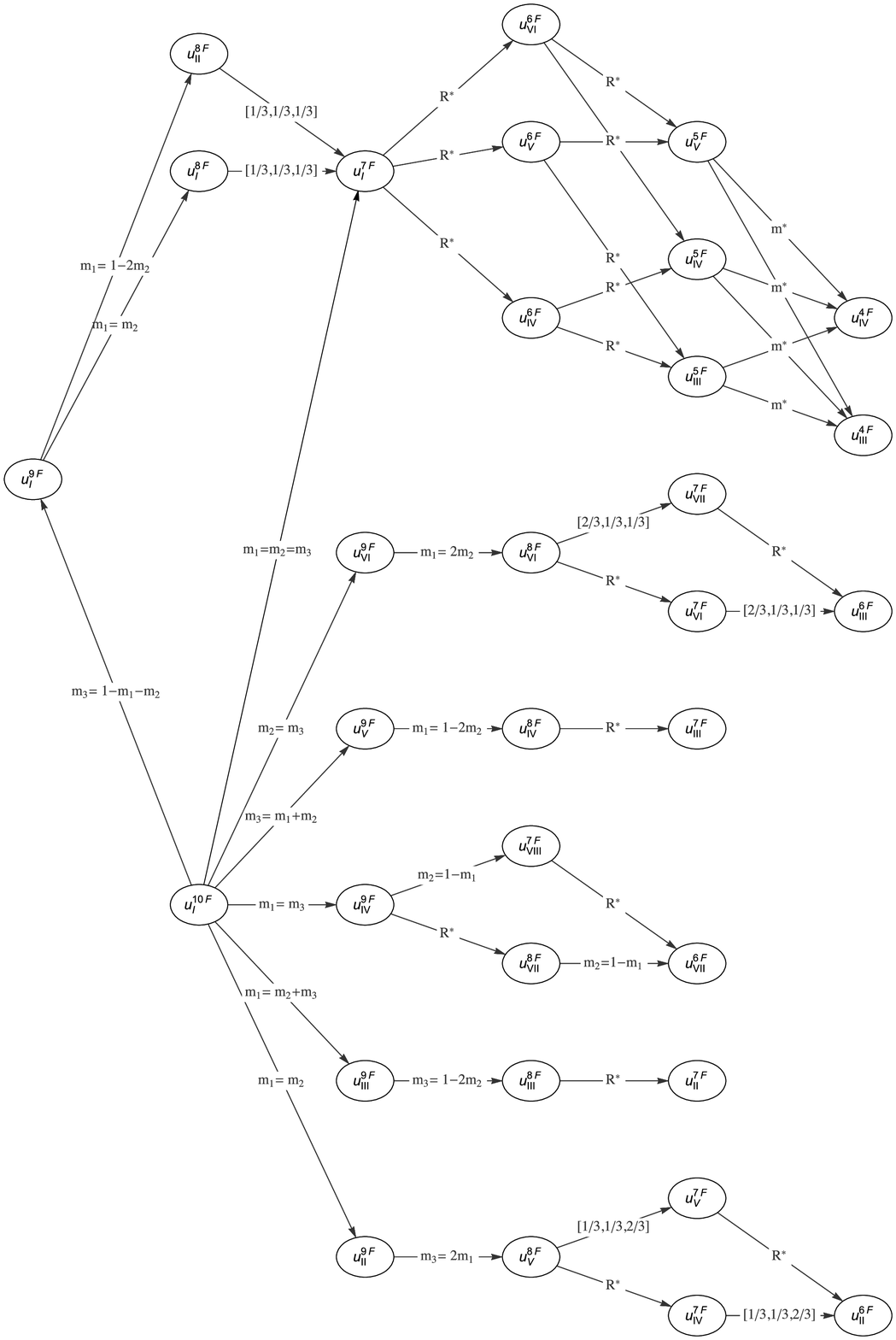}
\caption{\small A schematic representation of reductions of $u_I^{10F}(x)$ solution. Only most important reductions are shown and $N=2$ solutions are not included. The labels follow the structure of Fig.~\ref{fig:planes}. $R^{*}$ indicates that special choice of positions of domain walls is taken and $m^{*}$ indicates special choice of mass parameters described in the text.}
\label{fig:n3tree}
\end{figure}

The upper edge of the triangle on Fig.~\ref{fig:plane3} defined by relations $m_2 = 1-m$, $m_1 = m$ gives new 7F solution
\begin{align}
u_{VIII}^{7F}(x) & = 2\log\bigl(1+e^{R_1}e^{-m x}+e^{R_1+R_2}e^{- x}+e^{-(1+m) x}\bigr)\,, \\
\Omega_0 & = 1+2e^{R_1}e^{-m x}\bigl(1-m^2\bigr)+2me^{-(1+m)x}\Bigl(e^{2R_1+R_2}(2-m)-2-m\Bigr)
\nonumber \\ & +e^{2R_1}e^{-2m x}+e^{2R_1+2R_2}e^{-2x} +2e^{R_1+R_2}e^{-(2+m)x}\bigl(1-m^2\bigr)+e^{-2(1+m)x}\,.
\end{align}
We can reduce this solution further by setting $R_2 = -2R_1 +\log (2+m)/(2-m)$ to obtain a new 6F solution
\begin{align}
u_{VII}^{6F}(x) & = 2\log\bigl(1+e^{R_1}e^{-m x}+e^{-R_1}e^{- x}\tfrac{2+m}{2-m}+e^{-(1+m) x}\bigr)\,, \\
\Omega_0 & = 1+2e^{R_1}e^{-m x}\bigl(1-m^2\bigr)
\nonumber \\ & +e^{2R_1}e^{-2m x}+e^{-2R_1}e^{-2x}\tfrac{(2+m)^2}{(2-m)^2} +2e^{-R_1}e^{-(2+m)x}\bigl(1-m^2\bigr)\tfrac{2+m}{2-m}+e^{-2(1+m)x}\,,
\end{align}
which cannot be reduced further.
With this observation, we have exhausted all possible reductions in $N=3$ case.
We summarize the structure of reductions on Fig.~\ref{fig:n3tree}. 

\section{Conclusions}\label{sec:conclusions}

In this paper, we have shown many new exact solutions of the master equation \refer{eq:masterw}. Our main claim is that 
the class of functions
\begin{equation}\label{eq:class}
2\log\Bigl(1+\sum\limits_{i=1}^{N}\prod\limits_{j=1}^{i}e^{R_i-m
_i x}\Bigr) + 3\log\Bigl(1+\sum\limits_{i=1}^{\tilde N}\prod\limits_{j=1}^{i}e^{\tilde R_i-\tilde m
_i x}\Bigr)
\end{equation}
exhausts all exact solutions of the master equation for domain walls \refer{eq:masterw}, which can be written as a logarithm of the sum of exponentials. Of course, since we cannot prove this statement, we regard it as a conjecture.

In this paper we have studied two subgroups of the class \refer{eq:class} in detail. In the Sec.~\ref{sec:results} we have developed a concept of chains, originally spotted in \cite{Sakai3}. These are the solutions, which can be written as sum of single-wall solutions, or links, which we have denoted as $u_{II}^{3F}(x)$ and $u_{I}^{4F}(x)$ and their definitions can be found in Eq.~\refer{eq:link1} and Eq.~\refer{eq:link2}. We have discussed general properties of arbitrary long chains and derived general formulas for both the number of flavors of the minimal model in Eq.~\refer{eq:Nflavors} and the restriction on parameters in Eq.~\refer{eq:fullcond}, which must be satisfied in order to have well-defined moduli matrix.

In Sec.~\ref{sec:grinder} we have given a general discussion of solutions, which corresponds to taking $\tilde N =0$ limit in Eq.~\refer{eq:class}. In particular, we have shown that in terms of new variables $v(x) = e^{u(x)}-1$, these solutions can be understood as chains as well. More importantly, we have pointed out that solutions are related to each other via limits in their parameter space and thus form hierarchies. We showcased this hierarchy in detail for solutions up to $N_F=10$ flavors.

Throughout this paper, we argued that our exact solutions have two common features. First, they are all core-less and second, they are always non-elementary. 

Let us briefly discuss the first property. The lack of cores is intimately linked to the restriction on parameters in Eq.~\refer{eq:gencond}. Indeed, this  condition is in place to ensure that the lowest coefficient in $\Omega_0$ is never negative. At most, it can be zero, in which case remaining coefficients are positive. On the other hand, the core can develops only when all coefficients in $\Omega_0$ are very close to zero, since if $\Omega_0 =0$ the solution $u= g^2v^2 x^2$ is the unbroken phase inside the core. Given this observation, it is no surprise that our solutions cannot develop cores. 
Another, more heuristic reason, why our solutions are always core-less domain walls is that the functions we are working with, that is combinations of $\log(x)$ and $\exp(x)$, seems to be not sufficient to produce the shape of core-full domain walls, visualised in Fig.~\ref{fig:tensions}.

The fact that our solutions are non-elementary walls, is harder to understand.
More precisely, we claim that none of the multi-wall solutions of this paper have moduli, which could isolate elementary walls. For chains of Sec.~\ref{sec:results} this is easy to see because the single-wall links of which chains are made of $u_{II}^{3F}$ and $u_{I}^{4F}$ are non-elementary walls from the beginning. To confirm the same for the broadest class of solutions in Eq.~\refer{eq:class} is much harder to do in general. Intuitively, however, given that the constraint \refer{eq:gencond} makes any generic multi-wall configuration lighter than the lightest exact elementary wall $u_{I}^{2F}$, it is natural to conclude that our solutions are all non-elementary.
Despite this observation, the very fact that exact description of elementary walls seems to be much more difficult task than a description of compressed ones remains very puzzling.

In light of these findings, we are left with two challenges: first, is to find solutions for walls with well-developed cores and second, is to understand the disproportion between the number of exact elementary domain walls -- which are three: $u_{I}^{2F}$, $u_{I}^{2F}$ and $u_{III}^{3F}$ -- and the number of exact compressed domain walls -- which is potentially unlimited if the number of Higgs fields $N_F$ is arbitrarily large.

In closing, let us briefly comment about the impact of our findings to other topological solitons.

 The most direct implications can be made for composite solitons containing domain walls, such as wall-wall junctions and  wall-vortex junctions. The reason is that their master equations include Eq.~\refer{eq:masterw} as a subset, which hints that similar richness of exact solutions found for domain walls could be found there as well. 
Our preliminary results strongly favor this possibility and
we plan to explore this fully in a future study.

Another straightforward implication is that multi-flavor \emph{non}-Abelian domain walls (studied in \cite{Sakai2}) should possess a large number of exact solutions too. This can be seen easily by acknowledging the fact that solutions in Abelian model can be embedded into a non-Abelian case. More precisely, for a class of so-called $U(1)$-factorizable solutions, the non-Abelian analog of the master equation can be shown to decompose into a direct sum of independent Abelian master equations. Therefore any solution presented here can be used to construct a $U(1)$-factorizable solutions in a  non-Abelian model. The question whether our approach can help to find exact solutions which are not $U(1)$-factorizable provides another interesting direction for future work.
 
Lastly, an indirect implication of our results is that it raises expectations about number and abundance of exact solutions in \emph{any} master equation. In particular, in the paper \cite{Boojums} F.B. and others found several new exact solutions of semi-local vortices, hinting a very rich structure of exact solutions there. We plan to elaborate on these findings in the future.
 
\section*{Acknowledgment}

F. B. is indebted to Masato Arai and Minoru Eto for their help and countless consultations.
F. B. is an international research fellow of the Japan Society for the Promotion of Science.
This work was supported by Grant-in-Aid for JSPS Fellows, Grant Number 26004750.


\end{document}